\documentclass[prd,twocolumn,showpacs,preprintnumbers,nofootinbib,amsmath,eqsecnum]{revtex4}

\usepackage{epsfig}

\usepackage{graphicx}

\newcommand{\req}[1]{(\ref{#1})}
\newcommand{\nn}{\nonumber}
\newcommand{\be}{\begin{equation}}
\newcommand{\ee}{\end{equation}}
\newcommand{\ba}{\begin{eqnarray}}
\newcommand{\ea}{\end{eqnarray}}

\def\muF{\mu^2_F}
\def\muR{\mu^2_R}
\def\muO{\mu^2_0}
\def\als{\alpha_s}
\def\eps{\epsilon}
\def\veps{\varepsilon}
\def\qbq{q\overline{q}}
\def\ubu{{u}\overline{u}}
\def\dbd{{d}\overline{d}}
\def\sbs{{s}\overline{s}} 
\def\cbc{{c}\overline{c}}
\def\mev{~{\rm MeV}}
\def\gev{~{\rm GeV}}

\newcommand{\da}{{distribution amplitude}}
\newcommand{\ov}[1]{\overline#1} 
\newcommand{\tr}[1]{{\bf #1}_\perp}

\newcommand{\sla}{\hspace*{-0.20cm}/}

\begin{document}

\preprint{WU B 02-02, hep-ph/0210045}

\title{The two-gluon components of the $\eta$ and $\eta'$ mesons
 to leading-twist accuracy}

\author{Peter Kroll}
\email{kroll@physik.uni-wuppertal.de}
\author{Kornelija Passek-Kumeri\v{c}ki}
\email{passek@physik.uni-wuppertal.de }
\thanks{on leave of absence from the Rudjer Bo\v{s}kovi\'{c} Institute, 
Zagreb, Croatia .} 
\affiliation{Fachbereich Physik, Universit\"at Wuppertal,
        42097 Wuppertal, Germany}

\date{\today}

\begin{abstract}
We critically reexamine the formalism for treating the 
leading-twist contributions from the two-gluon Fock components
occurring in hard processes that involve $\eta$ and $\eta'$ mesons
and establish a consistent set of conventions for the definition of
the gluon \da{}, the anomalous dimensions as well as  the projector  
of a two-gluon state onto an $\eta$ or $\eta'$ state. We calculate 
the $\eta$, $\eta'$--photon transition form factor to order $\als$ 
and show  the cancellation of the collinear and UV singularities 
explicitly. An estimate of the lowest Gegenbauer coefficients of the 
gluon and quark distribution amplitudes is obtained from a fit to 
the $\eta$, $\eta'$--photon transition form factor data. 
In order to elucidate the role of the 
two-gluon Fock component further,
we analyze electroproduction of $\eta, \eta'$ mesons and the 
$g^*g^*\eta (\eta')$ vertex. 
\end{abstract}


\maketitle

\section{Introduction}
The description of hard exclusive processes involving light mesons is 
based on the factorization of the short- and long-distance dynamics
\cite{sHSA,LepageB80}. The former is represented by process-dependent,
perturbatively calculable parton-level subprocess amplitudes, in which 
the mesons are replaced by their valence Fock components, while the 
latter is described by process-independent meson \da s. This work is 
focused on hard reactions involving $\eta$ and $\eta'$ mesons. These 
particles as other flavor neutral mesons possess SU(3)$_{\rm F}$ 
singlet and octet valence Fock components and, additionally, 
two-gluon ones; to all three of them correspond \da s. This feature
leads, on the one hand, to the well-known {\it flavor mixing} which, 
for the $\eta-\eta'$ system, has been extensively studied (for a
recent review, see \cite{Feldmann99}) and, on the other hand, as a 
further complication, to {\it mixing} of the singlet and gluon \da s   
{\it under evolution}. On the strength of more and better experimental
data, the interest in hard reactions involving $\eta$ and $\eta'$
mesons and, consequently, in the role of the two-gluon Fock component,
has been renewed. Examples of such reactions are the meson-photon
transition form factors, photo- and electroproduction of mesons or
charmonium and $B$-meson decays. 

Mixing of the singlet and gluon \da s has been investigated in a
number of papers 
\cite{Terentev81}--\cite{BelitskyM98}.
Apart from differences in the notation and occasional misprints, 
different prefactors appear in the evolution kernels and in the
expressions for the anomalous dimensions. Often the full set of
conventions for kernels, anomalous dimensions, the
gluon \da{} and the gluon-meson projector is not provided and/or 
it is not easy to extract. This makes the comparison of the various
theoretical results and their applications difficult. We therefore
reexamine the treatment of the gluon \da{} and its mixing with the
singlet one. This analysis is performed in the context of the
$\eta\gamma$ and $\eta'\gamma$ transition form factors. Applying the
methods proposed in \cite{MelicNP01}, we calculate them to
leading-twist accuracy and include next-to-leading order (NLO)
perturbative QCD corrections. Our investigation enables us to
introduce and to test the conventions for the ingredients of a 
leading-twist calculation for any hard process that involves $\eta$ 
or $\eta'$ mesons. The most crucial test of the consistency of our 
set of conventions is the cancellation of the collinear singularities 
present in the parton-subprocess amplitude with the ultraviolet (UV) 
singularities appearing in the unrenormalized \da s. 
Our analysis permits a critical appraisal of the relevant literature 
\cite{Terentev81}--\cite{BelitskyM98}. 

In analogy with the analysis of the $\pi\gamma$ transition form factor
\cite{DiehlKV01}, we use our leading-twist NLO results for the
transition form factors to extract information on the $\eta$ and
$\eta'$ \da s from fits to the experimental data \cite{CLEO97,L398}. 
In order to make contact with experiment we have to adopt an appropriate 
$\eta - \eta'$ mixing scheme. We assume particle
independence of the \da s reducing so their number to three. Consequently,
flavor mixing is solely encoded in the decay constants
for which we use the values determined in \cite{FeldmannKS98}.

Our set of conventions, as abstracted from the calculation of the
transition form factor, is then appropriate for general use in
leading-twist calculations of hard exclusive reactions involving
$\eta$ and $\eta'$ mesons. We briefly discuss a few of them, namely,
electroproduction of the $\eta$ and $\eta'$ mesons and the vertex
$g^*g^*\eta (\eta')$, in order to learn more about the importance of
the gluon \da s. In contrast to the transition form factors, 
the two-gluon Fock components contribute in these reactions to the
same order of the strong coupling constant, $\als$, as the
quark-antiquark ones. The two-gluon components also contribute to 
the decays $\chi_{cJ} \to \eta\eta, \eta'\eta'$. The analysis of these
decays is however intricate since the next higher Fock state of the 
$\chi_{cJ}$, $\cbc g$ contributes to the same inverse power of the
relevant hard scale, the charm quark mass, as the $\cbc$ state and has
to be taken into account in a consistent analysis \cite{BolzKS96}. We
therefore refrain from analysing these decays here.

The plan of the paper is the following: The calculation of the
meson-photon transition form factors is presented in Sec.\
\ref{sec:singlet}. In Sec.\ {\ref{sec:mix}} we discuss $\eta-\eta'$ 
flavor mixing while Sec.\ \ref{sec:form} is devoted to a comparison 
with experiment and the extraction of the size of the lowest
Gegenbauer coefficients of the quark and gluon \da s. In Sec.\ 
\ref{sec:appl} we investigate the role of the gluon \da{} in other
hard reactions. The summary is presented in Sec.\ \ref{sec:summ}. The 
paper ends with three appendices in which we compile the definitions 
of quark and gluon \da s (App.\ A), calculational details for the 
transition form factors (App.\ B) and some properties of the evolution 
kernels (App.\ C).

\section{The $P\gamma$ transition form-factor}   
\label{sec:singlet}
\subsection{The flavor-singlet case}
\label{ss:fscase}
As the valence Fock components of the pseudoscalar mesons 
$P=\eta, \eta'$, we choose SU(3)$_{\rm F}$ singlet and octet
combinations of quark-antiquark states \footnote{This should not be
mixed up with the usual singlet and octet basis frequently used for
the description of $\eta-\eta'$ mixing. Our ansatz is completely general.}
\begin{eqnarray}
|\qbq_{1}\rangle &=& |(\ubu+\dbd+\sbs)/\sqrt{3} \rangle\,, 
\nn \\[0.2cm]
|\qbq_{8}\rangle &=& |(\ubu+\dbd-2\sbs)/\sqrt{6} \rangle\,, 
\label{eq:osbasis}
\end{eqnarray}
and the two-gluon state $|gg\rangle$ which also possess
flavor-singlet quantum numbers and contributes to leading twist. 
The corresponding \da s are denoted by $\phi_{P1,8,g}$;
their formal definitions are given in App.\ \ref{app:convdef}.
We emphasize that here, in this section, we do not make use 
of a flavor mixing scheme since the theoretical treatment of the
transition form factors is independent of it. As usual the decay 
constants, defined by the vacuum-meson matrix elements of 
flavor-singlet or octet weak axial vector currents ($i=1,8$)
\be
\langle 0 | J^i_{\mu 5}(0) | P(p) \rangle = i f_P^i\, p_\mu\,,
\label{eq:current}
\ee
or rather the factors $f_P^i/(2\sqrt{2N_c})$, are pulled out of the
\da{}s ($N_c$ being the number of colors). 
Hence, the quark \da s are normalized to unity at any scale $\mu^2$
\begin{equation}
\int_0^1 du\, \phi_{Pi}(u,\mu^2) =1 \, ,
\label{qda-norm}
\end{equation}
as follows from \req{eq:current} and \req{eq:melemq1u}.
From \req{eq:melemg1u} one has
\be
\int_0^1 du\, \phi_{Pg}(u,\mu^2) =0 \, .
\label{gda-nrom}
\ee
There is no natural way to normalize the gluon \da. Since the
flavor-singlet quark and gluon \da s mix under evolution while the 
flavor-octet one evolves independently with the hard scale, it is 
convenient to pull out of the gluon \da{} the same factor as for the 
flavor-singlet quark one. 

As usual we parameterize the $\gamma^*(q_1,\mu)\,\gamma(q_2,\nu) 
\rightarrow P(p)$ vertex as 
\begin{equation}
    \Gamma^{\mu}= i \, e^2 \; F_{P\gamma}(Q^{2}) 
         \; \veps^{\mu \nu \alpha \beta} \; \eps_{\nu}(q_2) 
                          \;q_{1\alpha} q_{2\beta}\, ,
\label{eq:Gmu}
\end{equation}
where  
$Q^2=-q_1^2 \geq 0$  is the momentum transfer,
and
$F_{P\gamma}(Q^2)$ denotes the $P\gamma$ transition form factor.
It can be represented  as a sum of the flavor-octet and the
flavor-singlet contributions  
\be
F_{P\gamma}(Q^2) = F_{P\gamma}^{\,8}(Q^2) + F_{P\gamma}^{\,1}(Q^2)\,,
\label{eq:ossum}
\ee
where the latter one includes the quark and the gluon part.  
The leading-twist singlet contribution to order $\als$ is unknown,
while the octet contribution is well-known to this order,
one only has to adapt the result for the $\pi\gamma$ transitions 
\cite{piontff} suitably. We therefore perform a 
detailed analysis of the singlet contribution along the lines of the
flavor-octet analysis presented in \cite{MelicNP01}.

\begin{figure}
\centerline{\includegraphics[height=4cm]{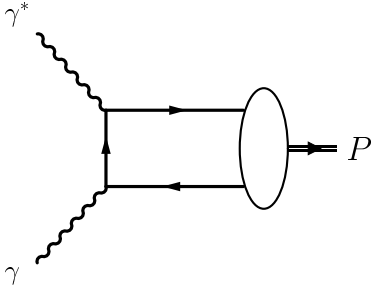}}
\caption{Lowest order Feynman diagram for the $\gamma^*\gamma\to P$
transition. A second diagram is obtained by interchanging the photon vertices.}
\label{fig:LOtff}
\end{figure}

For large  momentum transfer $Q^2$, the flavor-singlet 
contribution to the transition form factor can be represented as a 
convolution (see Fig.\ \ref{fig:LOtff} for a lowest order
Feynman diagram)  
\be
 F_{P\gamma}^{\,1}(Q^{2}) = 
  \frac{f_P^1}{2 \sqrt{2N_c}} \; 
    T^\dagger(u,Q^{2})   \, \otimes \,   \phi_P^{ur}(u)\,,
\label{eq:etffCF1}
\ee
where the symbol $\otimes$ represents the usual convolution 
$
A(z) \otimes B(z) = \int_0^1 dz A(z) B(z) \, .
$
We employ a two-component vector notation
\begin{equation}
\phi_P^{ur}(u) \equiv \left(
\begin{array}[c]{c}
\phi^{ur}_{Pq}(u) \\[0.2em]
\phi^{ur}_{Pg}(u) 
\end{array}
\right) \, ,\qquad
T(u,Q^{2}) \equiv\left(
\begin{array}[c]{c}
T_{\qbq}(u,Q^{2})\\[0.2em]
T_{gg}(u,Q^{2})
\end{array} \right)\,,
\label{eq:etaPhi}
\end{equation}
and switch to the more generic notation $\phi_{Pq}\equiv\phi_{P1}$. 
The unrenormalized quark and gluon distribution amplitudes 
$\phi^{ur}_{Pq}$ and $\phi^{ur}_{Pg}$ are defined in Eqs.\ 
(\ref{eq:unPhiqgen}) and (\ref{eq:unPhiggen}). 
The parton-level subprocesses amplitudes for 
$\gamma^* \gamma \rightarrow q \overline{q} $, and $\gamma^*
\gamma \rightarrow g g $ are denoted by
$T_{\qbq}$ and $T_{gg}$, respectively;
the Lorentz structure is factorized out as in \req{eq:Gmu}.  

The \da s $\phi^{ur}_{Pq}$ and $\phi^{ur}_{Pg}$ require
renormalization which introduces  mixing of the composite operators
$\bar{\Psi}(-z) \, \gamma^+ \gamma_5 \, \Omega \, \Psi(z)$
and $G^{+ \alpha}(-z) \, \Omega \, \widetilde{G}^{\: \: \: +}_{\alpha}(z)$.
The unrenormalized \da{}
$\phi^{ur}_P$ is related to the renormalized one,
$\phi_P$,  by
\begin{equation}
\phi^{ur}_P(u) =
 Z(u,x,\muF) \otimes \phi_P(x,\muF)
\, ,
\label{eq:etaPhiun}
\end{equation}
where the UV-divergent renormalization matrix takes the form 
\begin{equation}
Z\equiv
\left(
\begin{array}{cc}
Z_{qq} & Z_{qg} \\[0.2em]
Z_{gq} & Z_{gg}
\end{array}
\right)
\, .
\label{eq:Z}
\end{equation}
Here, $\muF$ represents the scale at which the singularities and, 
hence, soft and hard physics, are factorized.
Owing to the fact that quarks and gluons are taken to be 
massless and onshell, $T_{\qbq}$ and $T_{gg}$, calculated beyond
leading order, contain collinear singularities. The validity of
factorization into hard and soft physics, as expressed in
(\ref{eq:etffCF1}), requires the cancellation of these singularities
with the UV ones from the renormalization of the \da s. 
Hence, the hard scattering amplitude defined by
\be
T_{H}^\dagger(x,Q^{2},\muF) =  T^\dagger(u,Q^2) \otimes 
Z(u,x,\muF)\, ,
\label{eq:etaT}
\ee
must be finite. 
Below we explicitly show this cancellation to NLO.
Provided the cancellation of the singularities holds, the transition
form factor can be expressed in terms of finite hard scattering
and \da s
\begin{equation}
    F_{P\gamma}^{\,1}(Q^{2})= 
        \frac{f^1_P}{2 \sqrt{2N_c}} \;
    T_{H}(x,Q^{2},\muF)^{\dagger} \,\otimes \phi_P(x,\muF) \,. 
\label{eq:etffcf}
\end{equation}

\subsection{The NLO hard-scattering amplitude}

We now proceed to the NLO calculation. 
The renormalization matrix $Z$, 
can be shown to have the following form 
\be
Z= \mathbf{1} + \frac{\als(\muF)}{4 \pi}\, \frac1{\eps}\, V^{(1)}
   + {\cal O}(\als^2) \,,
\label{eq:Zexp}
\ee
if dimensional regularization ($D=4-2 \eps$) is employed. 
Here $\mathbf{1}$ denotes the unit $2\times2$ matrix (with
diagonal elements $\delta(x-u)$), and the coefficient
$V^{(1)}=V^{(1)}(x,u)$ is a matrix%
\footnote{
Since we are only interested in the $\als$ term,
we suppress the label $1$ in the matrix elements of $V^{(1)}$.}
\begin{equation}
V^{(1)} \equiv
\left(
\begin{array}{cc}
V_{qq} & V_{qg} \\[0.2cm]
V_{gq} & V_{gg}
\end{array}
\right)
\, .
\label{eq:V1}
\end{equation}
The amplitudes $T_{q \bar{q}}$ and $T_{gg}$ have  well-defined 
expansions in $\als$, and after coupling-constant renormalization,
which introduces the renormalization scale $\muR$, they read 
\begin{eqnarray}
T_{q \bar{q}}(u)& = &\frac{N_{q\bar{q}}}{Q^2} \left[ T_{q
             \bar{q}}^{(0)}(u) 
 \right. \nn  \\ & & \left. 
  +  \frac{\als(\muR)}{4 \pi}\, 
            C_F\, \left( \frac{\muR}{Q^2} \right)^{\eps} 
             T_{q \bar{q}}^{(1)}(u) 
 + {\cal O}(\als^2) \right]\,,
\nonumber \\[0.2cm]
T_{gg}(u) & = & \frac{N_{gg}}{Q^2}
  \left[ \frac{\als(\muR)}{4 \pi} 
   \left( \frac{\muR}{Q^2} \right)^{\eps} 
T_{gg}^{(1)}(u)
      + {\cal O}(\als^2) \right]
\, .
\nn \\ &  &
\label{eq:expT}
\end{eqnarray}
The normalization factors $N_{\qbq}$ and $N_{gg}$ in (\ref{eq:expT}) 
are given by 
\be
N_{q \bar{q}}= 2 \;\sqrt{2N_c} \; C_{1} \, ,
\qquad 
N_{gg} = \sqrt{n_f\, C_F}\, N_{\qbq} \, ,
\label{eq:Nqq}
\end{equation}
where the flavor factor $C_{1}$ takes into account the quark
content of the $\qbq_1$ combination. It reads 
(see (\ref{eq:ffactors}))
\be
{}C_{1}=\frac{e_u^2+e_d^2+e_s^2}{\sqrt{n_f}}
\, .
\label{eq:c1}
\ee
The number  of flavors in the $\qbq_1$ is denoted by $n_f$ and 
$C_F=(N_c^2-1)/(2 N_c)$ is the usual color factor. 
$e_a$ is the charge of quark $a$ in units of the positron charge $e$.

Inserting (\ref{eq:Zexp}) and (\ref{eq:expT}) into (\ref{eq:etaT}) 
and using \req{eq:chalphaS}, we obtain
\begin{eqnarray}
T_{H,q \bar{q}} & = &\frac{N_{q\bar{q}}}{Q^2}
   \left\{ T_{q \bar{q}}^{(0)} + 
    \frac{\als(\muR)}{4 \pi}
     \left[ C_F\, T_{q \bar{q}}^{(1)}
   \left( \frac{\muR}{Q^2} \right)^{\eps} 
 \right. \right. \nn  \\ & & \left. \left.
     + \frac{1}{\eps} T_{q \bar{q}}^{(0)} \otimes V_{qq} 
   \left( \frac{\muR}{\muF} \right)^{\eps} \, \right]
      + {\cal O}(\als^2) \right\}\,,\; \nonumber \\[0.2cm]
T_{H, gg} & = &\frac{N_{gg}}{Q^2} \left\{
    \frac{\als(\muR)}{4 \pi} 
  \left[ 
        T_{gg}^{(1)} 
   \left( \frac{\muR}{Q^2} \right)^{\eps} 
 \right. \right. \nn  \\ & & \left. \left.
   + \frac{N_{q\bar{q}}}{N_{gg}} \, 
  \frac{1}{\eps} T_{q \bar{q}}^{(0)} \otimes V_{qg}  
   \left( \frac{\muR}{\muF} \right)^{\eps} 
           \, \right]
      +{\cal O}(\als^2) \right\} .
  \nn  \\ & & 
\label{eq:expTH}
\end{eqnarray}
Results for $T_{\qbq}^{(0)}$, $T_{\qbq}^{(1)}$, $T_{gg}^{(1)}$,
and $V_{ij}$ and some details of their calculation are given in
App.\ \ref{app:calc}. Using the results for
$T^{(0)}_{\qbq}$ and $V_{qq}$, it is easy to verify that
\begin{equation}
  T^{(0)}_{\qbq}(u) \, \otimes \, V_{qq}(u,x)  = 
    C_F \; {\cal A}_{col,q \bar{q}}^{(1)}(x) \,,
\label{eq:11x1xVqq}
\end{equation}
with ${\cal A}_{col,q \bar{q}}^{(1)}$
being given in (\ref{eq:aqq}).
On the other hand, ${\cal A}_{col,q \bar{q}}^{(1)}$ 
is the residue of the $1/\eps$ pole in $T^{(1)}_{\qbq}$, 
see \req{eq:Tqq}. 
Hence, the collinear singularity present in $T_{\qbq}^{(1)}$ 
is canceled by the UV
singularity in $Z_{qq}$ and we arrive at a finite hard-scattering
amplitude for the $\gamma^* \gamma \to \qbq$ subprocess 
\begin{eqnarray}
\lefteqn{T_{H,\qbq}(x, Q^2,\muF)}
\nn \\  & = & \frac{N_{\qbq}}{Q^2}
\left\{ T_{H,\qbq}^{(0)}(x) 
+ \frac{\als(\muR)}{4 \pi} \,
C_F \; T_{H,\qbq}^{(1)}(x,Q^2,\muF) 
   \right. \nn  \\[0.1cm] & &  + {\cal O}(\als^2) \Big\}\,,
\label{eq:THqqexp}
\end{eqnarray}
where
\begin{eqnarray}
T_{H,\qbq}^{(0)}(x)&=&T_{\qbq}^{(0)}(x) \, ,
\nonumber \\
T_{H,q \bar{q}}^{(1)}(x,Q^2,\muF) &=& 
  -{\cal A}_{col,q \bar{q}}^{(1)}(x) \ln \frac{\muF}{Q^2}
  + {\cal A}_{q \bar{q}}^{(1)}(x) \,. 
\nn \\ & & 
\label{eq:THqq}
\end{eqnarray}
The quantities $T_{\qbq}^{(0)}$, ${\cal A}_{col,q \bar{q}}^{(1)}$, 
and ${\cal A}_{q \bar{q}}^{(1)}$ are given in
(\ref{eq:Tqq}, \ref{eq:aqq}).

Next, from \req{eq:Tqq} and \req{eq:Vqg}, we obtain
\begin{equation}
 T^{(0)}_{\qbq}(u) \, \otimes \, V_{qg}(u,x)  =
            \frac{N_{gg}}{N_{\qbq}} \,  
            {\cal A}_{col,gg}^{(1)}(x) \, ,
\label{eq:11x1xVqg}
\end{equation}
with ${\cal A}_{col,gg}^{(1)}$  defined in (\ref{eq:agg}).
Inserting this result into \req{eq:expTH} and taking into account
\req{eq:Tgg1}, we observe the cancellation of the collinear 
singularity present in $T_{gg}^{(1)}$ with the UV singularity of 
$Z_{qg}$, and we get the finite hard-scattering amplitude for the 
$\gamma^* \gamma \to gg$ subprocess  
\begin{equation}
T_{H,gg}(x, Q^2) =\frac{N_{gg}}{Q^2} 
                    \left[ \frac{\als (\muR)}{4 \pi}
        T_{H,gg}^{(1)}(x,Q^2,\muF) + {\cal O}(\als^2) \right]
   \, ,
\label{eq:THggexp}
\end{equation}
where $T_{H,gg}^{(1)}$ reads
\begin{equation}
T_{H,gg}^{(1)}(x,Q^2,\muF) = 
 - {\cal A}_{col,gg}^{(1)}(x) \ln \frac{\muF}{Q^2} +
 {\cal A}_{gg}^{(1)}(x)
  \, .
\label{eq:THgg1}
\end{equation}
The functions ${\cal A}_{col,gg}^{(1)}$ and ${\cal A}_{gg}^{(1)}$ 
are supplied in (\ref{eq:agg}).
\subsection{Evolution of the flavor-singlet quark and gluon \da s}
\label{s:DAev}
We now turn to the discussion of the distribution amplitude
$\phi_P$ and its evolution.
The matrix $Z$ is related to the evolution of the
distribution amplitude, and $V^{(1)}$ in \req{eq:Zexp} represents 
the kernel which governs the leading-order (LO) evolution of 
the flavor-singlet 
\da. By differentiating (\ref{eq:etaPhiun}) with respect to $\muF$ 
one obtains the evolution equation \cite{Terentev81,BaierG81} 
\begin{equation}
  \muF \frac{\partial}{\partial \muF} \phi_P(x,\muF)   =
   V(x,u,\als(\muF)) \, \otimes \, \phi_P(u,\muF)
         \, ,
\label{eq:eveq}
\end{equation}
where the evolution kernel $V$ reads  
\begin{equation}
   V  =  -Z^{-1} \, \otimes \, \left( \muF
       \frac{\partial}{\partial \muF} Z
              \right)\, .
\label{eq:VZ}
\end{equation}
We note in passing that the evolution equation
would have a more complicated form if the factor 
$f_{P1}/(2 \sqrt{2 N_c})$ was not pulled 
out of the gluon distribution amplitude.
Inserting (\ref{eq:Zexp}) into (\ref{eq:VZ}), and using
\req{eq:beta}, one easily sees that
\be
V=\frac{\als(\muF)}{4\pi}\, V^{(1)} + {\cal O}(\als^2)
\, .
\ee

The results for the LO kernel $V^{(1)}$ are given in (\ref{eq:Vqq})
and (\ref{eq:Vqg}-\ref{eq:Vgg}). The anomalous dimensions that 
control the evolution of the \da s can be read off from the 
relations \req{eq:VC}:
\begin{eqnarray}
\gamma^{\,qq}_n &=& C_F \left[
   3 + \frac{2}{(n+1)(n+2)} - 4 \sum_{i=1}^{n+1} \frac{1}{i} \right]
                  \, , \nn \\[0.3em]
\gamma^{\,qg}_n &=& \sqrt{n_f C_F} \; 
\              \frac{n (n+3)}{3 (n+1) (n+2)} \qquad n\ge2
                        \, , \nn \\[0.3em]
\gamma^{\,gq}_n &=& \sqrt{n_f C_F} \;  
                     \frac{12}{\phantom{3} (n+1) (n+2)} \qquad n\ge2
                      \, , \nn \\[0.3em]
\gamma^{\,gg}_n &=& \beta_0 + N_c \left[ \frac{8}{(n+1)(n+2)} 
             - 4 \sum_{i=1}^{n+1} \frac{1}{i} \right] \qquad n\ge2
                       \, .  
\nn \\ & &
\label{eq:andim}
\end{eqnarray}

To leading order in $\als$ the evolution equation \req{eq:eveq} can be
solved by diagonalizing the kernel $V$ or rather the matrix of the 
anomalous dimensions. The eigenfunctions can be expanded upon the
Gegenbauer polynomials $C_n^{m/2}$ with coefficients $B_{Pn}^{(\pm)}$ 
which evolve with the eigenvalues $\gamma_n^{(\pm)}$ of the matrix of
the anomalous dimensions 
\begin{equation}
\gamma^{\,(\pm)}_n = \frac{1}{2} \left[ \gamma^{qq}_n+ \gamma^{gg}_n 
        \pm \sqrt{ \left( \gamma^{qq}_n- \gamma^{gg}_n \right)^2
        +4 \gamma^{qg}_n \gamma^{gq}_n } \right] \, .
\label{eq:gamma+-}
\end{equation}
The two components of the \da{} $\phi_P$ possess the expansion
\begin{eqnarray}
\phi_{Pq}(x,\muF)  &=& 6 x\, (1-x)\, \Big[ 1 
\nn \\ & &
     + \sum_{n=2, 4, \ldots}
    {\  B_{Pn}^{\,q}(\muF)} \:  C_n^{\,3/2}(2 x -1) \Big]\,,
                                              \nn\\[0.2cm] 
\phi_{Pg}(x,\muF)  &=&  x^2 (1-x)^2 
\nn \\ & & \times
     \sum_{n=2, 4, \ldots}
              { B_{Pn}^{\,g}(\muF)} \:  C_{n-1}^{\,5/2}(2 x -1) \, ,
\nn \\ & &
\label{eq:solphi} 
\end{eqnarray}
where only the terms for even $n$ occur as a consequence of
\req{eq:DAsim}. The expansion coefficients in
\req{eq:solphi} are related to those of the eigenfunctions by
\begin{eqnarray}
{ B_{Pn}^{\,q}\,(\muF)} & = & B_{Pn}^{(+)}(\mu_0^2) 
         \left( \frac{\als(\mu_0^2)}{\als(\muF)} \right)^{%
         { \gamma^{\,(+)}_n}/{\beta_0}} 
\nn \\ & &
     + \, 
         { \rho_n^{\,(-)}} \,  B_{Pn}^{(-)}\,(\mu_0^2) 
         \left( \frac{\als(\mu_0^2)}{\als(\muF)} \right)^{%
         { \gamma^{\,(-)}_n}/{\beta_0}}\hspace*{-1mm} ,
                                         \nn\\[0.2cm]
{ B_{Pn}^{\,g}\,(\muF)} & = & {\rho_n^{\,(+)}} \, 
               B_{Pn}^{(+)}\,(\mu_0^2) 
               \left( \frac{\als(\mu_0^2)}{\als(\muF)} \right)^{%
               {\gamma^{\,(+)}_n}/{\beta_0}} 
\nn \\ & &
     +  \, B_{Pn}^{(-)}\,(\mu_0^2) 
               \left( \frac{\als(\mu_0^2)}{\als(\muF)} \right)^{%
               {\gamma^{\,(-)}_n}/{\beta_0}}\hspace*{-1mm}. 
\label{eq:Bng}
\end{eqnarray}
The coefficients $B_{Pn}^{(\pm)}\,(\mu_0^2)$ respective
$B_{Pn}^{\,q,g}\,(\mu_0^2)$, where $\mu_0^2$ is the initial scale of
the evolution, 
represent the non-perturbative  input to a calculation of the transition
form factors and are, at present, not calculable with a sufficient
degree of accuracy. The parameters $\rho_n^{\,(\pm)}$ read 
\begin{equation}
\rho_{n}^{\,(+)} =  6 \; 
        \frac{\gamma^{\,gq}_n}{\gamma^{\,(+)}_n - \gamma^{\,gg}_n}\,,
        \qquad
\rho_{n}^{\,(-)} = \frac{1}{6}\;  
        \frac{\gamma^{\,qg}_n}{\gamma^{\,(-)}_n - \gamma^{\,qq}_n}\, . 
\label{eq:rho} 
\end{equation}

We note that the anomalous dimensions satisfy the relation
\begin{equation}
\frac{\gamma^{\,qg}_n}{\gamma^{\,(\pm)}_n - \gamma^{\,qq}_n} =
\frac{\gamma^{\,(\pm)}_n - \gamma^{\,gg}_n}{\gamma^{\,gq}_n}\, .
\end{equation}
Comparison of \req{eq:andim} and (\ref{eq:gamma+-})
reveals that  $ \gamma_n^{(+)}\approx \gamma^{qq}_n$
for all $n$
and $\gamma_n^{(+)} \to \gamma^{qq}_n$ for $n \to \infty$.

It is important to realize that any change of the definition of 
the gluon \da{} \req{eq:unPhiggen} is 
accompanied by a corresponding change in the hard scattering 
amplitude. Suppose we change $\phi_{Pg}$ by a
factor $\sigma$ 
\be
    \phi^{\,\sigma}_{Pg} = \sigma\, \phi_{Pg}\,.
\label{eq:phigsigma}
\ee 
Since any physical quantity, as for instance the transition form
factor, must be independent of the choice of the convention, 
the projection \req{eq:gg} of $gg$ state onto a pseudoscalar meson
state is to be modified by a factor $1/\sigma$, i.e.,
\begin{equation}
{\cal P}^{g\,\sigma}_{\,\mu \nu} =
   \frac{1}{\sigma}\, {\cal P}^{g}_{ \, \mu \nu}\,,
\label{eq:ggsigma}
\end{equation}
and the hard-scattering amplitude becomes altered accordingly.
As an inspection of Eqs.\ (\ref{eq:solphi})-(\ref{eq:rho})
reveals, the change of the definition of the gluon
\da{} \req{eq:phigsigma} 
has to be converted into a change of the off-diagonal anomalous
dimensions and the Gegenbauer coefficients $B_{Pn}^{(\pm)}$ in order
to leave the quark \da{} as it is:
\begin{equation}
\gamma^{\,qg, \sigma}_n = \frac{1}{\sigma} \, \gamma^{\,qg}_n\,,
 \qquad
\gamma^{\,gq, \sigma}_n = \sigma \, \gamma^{\,gq}_n\,,
\label{eq:gammasigma}
\end{equation}
and
\begin{equation}
{B}^{(-)\,\sigma}_{Pn}\,(\mu_0^2)=\sigma B^{(-)}_{Pn}\,(\mu_0^2)\,,
\qquad 
B^{(+)\,\sigma}_{Pn}\,(\mu_0^2)= B^{(+)}_{Pn}\,(\mu_0^2)\,,
\label{eq:Bsigma}
\end{equation}
implying
\be
B^{\,g\,\sigma}_{Pn}\,(\muF)=\sigma\, B^{\,g}_{Pn}\,(\muF)\,,\qquad
B^{\,q\,\sigma}_{Pn}\,(\muF)= B^{\,q}_{Pn}\,(\muF)\,.
\label{eq:BsigmaG}
\ee
We finally mention that,
as can be easily seen from Eq. \req{eq:phigsigma}
and the evolution equation \req{eq:eveq},
along with the change of the anomalous dimensions
\req{eq:gammasigma} the kernels $V_{qg}$ and $V_{gq}$ become
modified. 

The results for the anomalous dimensions can also be understood in the 
operator language, i.e., by considering the impact of a change of 
the definition of the gluonic composite operator on the anomalous dimensions
(for comments on the use of the operator product expansion, see, for
instance, Refs. \cite{ShifmanV81,BaierG81}).   
One finds that only the anomalous dimensions
$\gamma^{\,qg}_n$ and $\gamma^{\,gq}_n$ become modified, while
the diagonal ones and the product
$\gamma^{\,qg}_n \, \gamma^{\,gq}_n$, and consequently 
the eigenvalues $\gamma_n^{(\pm)}$, remain unchanged.
Redefinition of the gluonic composite operator implies a corresponding
change of the gluon \da{}.

We are now in the position to compare the results presented in
this work with other calculations to be found in the literature.
The entire set of conventions is not always easy to extract from 
the literature since often only certain aspects of the 
flavor-singlet system are discussed. For instance, in Ref.\
\cite{BlumleinGR97} only the evolution kernels are investigated,
or in Ref.\ \cite{ShifmanV81} only the anomalous dimensions.
Using results from such work in a calculation of a hard process 
necessitates the use of corresponding conventions for the other 
quantities. Care is also required if elsewhere determined numerical 
results for the Gegenbauer coefficients $B_{Pn}^{(-)}$ or $B_{Pn}^{g}$
are employed since, according to \req{eq:Bsigma} and \req{eq:BsigmaG},
they are convention dependent.     
For future reference, we systematize in Tab.\ \ref{t:conventions} 
the important ingredients for the three conventions encountered 
in the literature. Our expressions for the kernels and the 
anomalous dimensions correspond to the ones obtained in 
\cite{Terentev81} (up to a typo in $V_{gg}$). 
In Refs.\ \cite{BelitskyM98etc,BelitskyM98} 
the anomalous dimensions controlling the
evolution of the forward and non-forward parton distribution 
were studied to NLO.
Since the non-diagonal anomalous dimension 
for the odd parity case coincides with our ones \cite{MikhailovR}, we 
observe that the convention $\sigma=\sqrt{C_F/n_f}$ is used in 
\cite{BelitskyM98etc,BelitskyM98}. 
The only result we do
not understand is the one presented in Ref.\ \cite{Ohrndorf81}: 
There is an extra factor of $1/2$ in $V_{gq}$ which changes the product of 
prefactors. Moreover, there are factors $1/3$ and $3$ apparently
missing in 
$\gamma_{qg}$ and $\gamma_{gq}$. We note that occasionally the factor 
$[x(1-x)]^{-1}$ appearing in our projector (\ref{eq:gg}) is absorbed into 
the gluon \da{} \cite{Ohrndorf81,BaierG81}. This arrangement is
accompanied by corresponding changes of the evolution kernels, see
\req{eq:altV}.    
\renewcommand{\arraystretch}{2}
\begin{table}
\begin{center}
\caption{List of common conventions for the anomalous dimensions and
the $gg$ projector. Quoted are the prefactors of the non-diagonal
anomalous dimensions (\ref{eq:andim}) and of the $gg$ projector
(\ref{eq:gg}) for various choices of $\sigma$ in (\ref{eq:ggsigma},
\ref{eq:gammasigma}). We also list references 
where these conventions
for the anomalous dimensions are used.}    
\vspace*{5mm}

\begin{tabular}{c||c|c|c||c}
$\sigma$ & 
$\gamma^{\,qg,\sigma}_n$ &
$\gamma^{\,gq,\sigma}_n$ &
${\cal P}^{g\sigma}_{\mu \nu}$ &
references \\ \hline
$1$ &
$\sqrt{n_f \, C_F}$ &
$\sqrt{n_f \, C_F}$ &
$1$ & \cite{Terentev81}\\
$\displaystyle \sqrt{\frac{n_f}{C_F}}$ & 
$C_F$ &
$n_f $ &
$\displaystyle \sqrt{\frac{C_F}{n_f}}$ &
\cite{ShifmanV81,BaierG81}\\
$\displaystyle \sqrt{\frac{C_F}{n_f}}$ &
$n_f $ &
$C_F$ &
$\displaystyle \sqrt{\frac{n_f}{C_F}}$ & 
\cite{BlumleinGR97,BelitskyM98}\\ \hline
\end{tabular}
\label{t:conventions}
\end{center}
\end{table}
\renewcommand{\arraystretch}{1}

Although, from the point of view of derivation, 
the conventions which lead to (\ref{eq:andim}) and (\ref{eq:gg}) 
seem to be the most natural ones,
it is perhaps more expedient to use the same conventions
for the anomalous dimensions as for polarized deep inelastic
lepton-proton scattering \cite{AhmedR76}, which correspond to 
\begin{equation}
       \sigma= \sqrt{\frac{n_f}{C_F}}
\, .
\label{eq:newconv}
\end{equation}
The corresponding set of conventions will be used in the rest of the
paper. The non-diagonal anomalous dimensions then read 
\begin{eqnarray}
\gamma^{\,qg}_n &\longrightarrow& C_F \; 
\              \frac{n (n+3)}{3 (n+1) (n+2)} \qquad n\ge2
                        \, , \nn \\[0.3em]
\gamma^{\,gq}_n &\longrightarrow& n_f \;  
                     \frac{12}{\phantom{3} (n+1) (n+2)} \qquad n\ge2
                      \, , 
\label{eq:andimNEW}
\end{eqnarray}
and the gluonic projector 
\be
{\cal P}^{g}_{\mu \nu, a b} \longrightarrow  \, \frac{i}{2}  \;  
        \sqrt{\frac{C_F}{n_f}} \; 
            \frac{\delta_{ab}}{\sqrt{N_c^2-1}}
         \; \frac{\veps_{\perp \mu \nu}}{u(1-u)} \,.
\label{eq:ggNEW}
\ee
Along with these definitions, Eqs. (\ref{eq:solphi})-(\ref{eq:rho}) 
have to be used. 

To the order we are working, the NLO evolution of the quark 
distribution amplitudes should in principle be included (the
convolution of the NLO term for $\phi_{Pg}$ with $T_{H,gg}$
contributes to order $\alpha_s^2$).
To NLO accuracy the Gegenbauer polynomials $C_n^{3/2}$ are no longer
eigenfunctions of the evolution kernel, so that their coefficients
$B_{Pn}^{i}$ do not evolve independently
\cite{BelitskyM98,Muller95}. In analogy with the pion case 
\cite{KrollR96}, the impact of the NLO evolution on the transition 
form factors is expected to be small compared with the NLO corrections
to the subprocess amplitudes. Therefore we refrain from considering
NLO evolution. 

\subsection{The NLO result for the transition form factor}
\label{ss:results}
To end this section we quote our final result for the flavor-singlet  
contribution to the $P\gamma$ transition form factor to 
leading-twist accuracy and NLO in $\als$. The result, obtained by inserting
\req{eq:THqqexp} and \req{eq:THggexp}
(multiplied by $\sigma^{-1}=\sqrt{C_F/n_f}$
according to the new normalization of the gluonic
projector) 
into \req{eq:etffcf}, is 
\begin{eqnarray}
\lefteqn{F_{P\gamma}^{\,1} (Q^{2}) } \nonumber \\
&=& \frac{f^1_P \, C_{1}}{Q^2} \;
               \left\{ T_{H,q \bar{q}}^{(0)}(x) \otimes 
               \phi_{Pq} (x,\muF) \right. \nn \\ 
           & & \left. + \frac{\als(\muR)}{4 \pi} \, C_F \, \left[
               \;T_{H,\qbq}^{(1)}(x,Q^2,\muF) \otimes 
            \phi_{Pq} (x,\muF) \right. \right. \nn \\ 
            & & \left. \left. \hspace*{2.2cm}+
                T_{H,gg}^{(1)}(x,Q^2,\muF) \otimes 
                \phi_{Pg}(x,\muF) \right]\,\right\}\,.
\nn \\ & &
\label{eq:Resetatff}
\end{eqnarray}
A subtlety has to be mentioned. The singlet decay constant, $f_P^1$, 
depends on the scale but the anomalous dimension controlling it is 
of order $\als^2$ \cite{Leutwyler97etc}. 
In our NLO calculation this effect is tiny and is to
be neglected as the NLO evolution of the \da{}.

For completeness and for later use we also quote the result for 
the flavor-octet contribution to the $P\gamma$ transition form 
factor at the same level of theoretical accuracy.
In our notation it reads
\begin{eqnarray}
\lefteqn{F_{P\gamma}^{\,8}(Q^{2})}
\nn \\ &=& \frac{f_P^{8} \, C_{8}}{Q^2} \;
       \left\{ T_{H,\qbq}^{(0)}(x) \otimes \phi_{P8}(x,\muF)
                               \right. \nonumber \\ 
          & & \left. + \frac{\als(\muR)}{4 \pi} \,  
        C_F \;T_{H,\qbq}^{(1)}(x,Q^2,\muF) \otimes \phi_{P8}(x,\muF) \:
                     \right\}  \, ,
\nn \\ & &
\label{eq:tffeta8}
\end{eqnarray}
where the renormalized hard scattering amplitude is given in
(\ref{eq:THqqexp}) and the charge factor $C_8$ is obtained with the
help of (\ref{eq:ffactors})
\be 
C_{8} =\frac{e_u^2+e_d^2-2 e_s^2}{\sqrt{6}} \, .
\label{eq:c8}
\ee
The octet distribution amplitude, $\phi_{P8}$, being fully  
analogous to the pion case, has the expansion
\begin{eqnarray}
\phi_{P8} (x,\muF)  &=&  6 x (1-x) \left[ 1 
  \right.  \nn \\ & & \left.
        + 
   \sum_{n=2, 4, \ldots} B_{Pn}^{\,8}(\muF) \; 
          C_{n}^{\,3/2}(2 x -1) \right]\, , 
\nn \\ & &
\label{eq:phi8}
\end{eqnarray}
where the Gegenbauer coefficients evolve according to \cite{sHSA}
\begin{equation}
B_{Pn}^{\,8}(\muF) = B_{Pn}^{\,8}(\mu_0^2) \, 
               \left( \frac{\als(\mu_0^2)}{\als(\muF)} 
               \right)^{\gamma_n^{\,qq}/\beta_0} \, .
\label{eq:Bn8muF}
\end{equation}
Summing the flavor-singlet and octet contributions according to
\req{eq:ossum}, we arrive at the full transition form factors 
for the physical mesons.

As has been pointed in Refs.\ \cite{Feldmann99,DiehlKV01,FeldmannK97},
in the limit $Q^2\to\infty$ where the quark \da{}s
evolve into the asymptotic form 
\be
\phi_{AS}(x)=6x(1-x) 
\label{eq:phiAS}
\ee
and the gluon one to zero, the transition form factor becomes
\be
F_{P\gamma} \stackrel{Q^2\to\infty}{\longrightarrow} \frac{\sqrt{2}
                   \,{f}_P^{\rm eff}}{Q^2} \left[1 - 
                             \frac53 \frac{\als}{\pi}\right]\,.
\label{f-asy}
\ee
${f}_P^{\rm eff}$ combines the decay constants with the 
charge factors $C_i$
\be
{f}_P^{\rm eff} = \frac{1}{\sqrt{3}} \left[f_P^8 
              + 2 \sqrt{2} f_P^1 \right] \,. 
\ee
The result \req{f-asy} holds also for the case of the pion with
${f}_\eta^{\rm eff}$ replaced by $f_\pi$. In \cite{Feldmann99} 
an interesting observation has been reported: if the transition 
form factors for the $\pi$, $\eta$ and $\eta'$ are scaled by 
their respective asymptotic results, the data for these 
processes \cite{CLEO97,L398} fall on top of each other 
within experimental errors. This can be regarded as a hint at 
rather similar forms of the quark \da s in the three cases and 
a not excessively large gluon one.

\section{$\eta$--$\eta'$ mixing}
\label{sec:mix}
Using the results \req{eq:Resetatff} and \req{eq:tffeta8}
for the transition form factors, one may analyze the experimental 
data obtained by CLEO \cite{CLEO97} and L3 \cite{L398} with the aim 
of extracting information on
the six distribution amplitudes $\phi_{Pi}(x,\muO)$, $i=1,8,g$
or rather on their lowest Gegenbauer coefficients $B_{Pn}^i(\muO)$.
In principle, this is an extremely interesting program since it 
would allow for an investigation of $\eta-\eta'$ flavor mixing
at the level of the \da s. In practice, however, this program is
to ambitious since the present quality of the data is insufficient
to fix a minimum number of six coefficients which occur if the
Gegenbauer series is truncated at $n=2$. Thus, we are forced to
change the strategy and to employ a flavor mixing scheme right from
the beginning in order to reduce the number of free parameters.
 
Since in hard processes only small spatial quark-antiquark 
separations are of relevance, it is sufficiently suggestive to 
embed the particle dependence and the mixing behaviour of the 
valence Fock components solely into the decay constants, which 
play the role of wave functions at the origin. Hence, following
\cite{FeldmannKS98,FeldmannK97}, we take 
\be
     \phi_{P i} = \phi_i\,,
\label{new-ansatz}
\ee  
for $i=8,1,g$. This assumption is further supported by the 
observation \cite{FeldmannK97,JKR} that, as for the 
case of the pion \cite{DiehlKV01,KrollR96,MusatovR97},
the quark \da s for the $\eta$ and $\eta'$ mesons seem to be close 
to the asymptotic form $\phi_{AS}(x)$ for which the particle 
independence (\ref{new-ansatz}) holds trivially . 
Note that we switch now back to the original notation for the 
singlet \da{} introduced in Sec.\ \ref{ss:fscase}:
\be
\phi_{P1}\equiv \phi_{Pq}\,, \qquad  B_{Pn}^1 \equiv B_{Pn}^q\,.
\ee

The decay constants can be parameterized as
\cite{FeldmannKS98,Leutwyler97etc}
\ba
f_{\eta\phantom{'}}^8 &=& f_8 \cos{\theta_8}\,, 
        \qquad f_{\eta\phantom{'}}^1 = -f_1 \sin{\theta_1}\,, \nn\\
f_{\eta'}^8 &=& f_8 \sin{\theta_8}\,, 
              \qquad f_{\eta'}^1 =\phantom{-}f_1 \cos{\theta_1}\,.
\label{mix81}
\ea
Numerical values for the mixing parameters  have been
determined on the basis of the quark-flavor mixing scheme 
\cite {FeldmannKS98}  :
\ba
f_8 &=& 1.26 f_\pi\,, \qquad \theta_8 = -21.2^\circ\,, \nn\\
f_1 &=& 1.17 f_\pi\,,  \qquad \theta_1=-9.2^\circ\,.
\label{eq:num81}
\ea
The value of the pion decay constant is $f_\pi=0.131 \gev$.
As observed in \cite{FeldmannKS98} (see also 
\cite{Feldmann99}) $\eta-\eta'$ flavor mixing can 
be parameterized in the simplest way in the quark-flavor basis.
The mixing behaviour of the decay constants in that basis follows 
the pattern of state mixing, i.e. there is only one mixing angle.
The basis states of the quark-flavor mixing scheme are defined by
\begin{eqnarray}
|\eta_q\rangle &=& \cos{\varphi}\, |\eta\rangle + 
                       \sin{\varphi}\, |\eta'\rangle, 
\nn \\ 
|\eta_s\rangle&=& -\sin{\varphi}\, |\eta\rangle + 
                    \cos{\varphi}\, |\eta'\rangle ,
\label{qsbasis}
\end{eqnarray}   
and the strange and non-strange decay constants are assumed to mix as
\ba
f_{\eta\phantom{'}}^q &=& f_q \cos{\varphi}\,, \qquad 
            f_{\eta\phantom{'}}^s = -f_s \sin{\varphi}\,, \nn\\
f_{\eta'}^q &=& f_q \sin{\varphi}\,, \qquad 
                f_{\eta'}^s =\phantom{-}f_s \cos{\varphi}\,.
\label{mixqs}
\ea
As demonstrated in \cite{FeldmannKS98} this ansatz 
is well in agreement with experiment. The occurrence of only one 
mixing angle in this scheme is a consequence of the smallness of 
OZI rule violations which amount to only a few percent and can 
safely be neglected in most cases. $SU(3)_F$ symmetry, on the 
other hand, is broken at the level of $10 -20\%$ as can be 
seen, for instance, from the values of the decay constants $f_8$ 
and $f_1$, and cannot be ignored. 

Using \req{eq:osbasis} and particle independence, we obtain for the 
valence Fock components of the basis states \req{qsbasis}
\ba
|\eta_q\rangle &=& \frac{f_q}{2\sqrt{2N_c}}\, 
               \left[\phi_q(x,\muF)\, |\qbq\,\rangle
               +      \phi_{\rm opp}(x,\muF)\, |\sbs\,\rangle
\right. \nn \\ & & \left.
               + \sqrt{2/3}\; \phi_g(x,\muF)\, |gg\rangle \right]
                                            \,, \nn\\[0.2em]  
|\eta_s\rangle &=& \frac{f_s}{2\sqrt{2N_c}}\,
               \left[ \phi_{\rm opp}(x,\muF)\, |\qbq\,\rangle 
                 +  \phi_s(x,\muF)\,|\,\sbs\,\rangle 
\right. \nn \\ & & \left.
              \, + \phi_g(x,\muF)\, |gg\rangle/ \sqrt{3}\right]
                        \,,
\label{qsfock} 
\ea
where $\qbq$ is short for the combination $(\ubu+\dbd)/\sqrt{2}$ and
\be
\begin{array}{ccc}
\phi_q=\frac13 (\phi_8 + 2\phi_1)\,, & \quad &
\phi_s=\frac13 (2\phi_8 + \phi_1)\,, \\[0.15cm]
\multicolumn{3}{c}{\displaystyle
\phi_{\rm opp} = \frac{\sqrt{2}}{3} (\phi_1 - \phi_8)\,.
}
\end{array}
\ee
In deriving \req{qsfock} we made use of the relations
\ba
\cos(\varphi-\theta_8) &=& \frac{1}{\sqrt{3}} \frac{f_q}{f_8}
                                                    \,, \qquad 
  \cos(\varphi-\theta_1) = \frac{1}{\sqrt{3}} \frac{f_s}{f_1}\,,\nn\\  
\sin(\varphi-\theta_8) &=& \sqrt{\frac{2}{3}} \frac{f_s}{f_8}
                                                     \,, \qquad 
  \sin(\varphi-\theta_1) = \sqrt{\frac{2}{3}} \frac{f_q}{f_1}\,,
\nn \\ & &
\label{trig}
\ea
which can readily be obtained from results on decay constants and 
mixing angles reported in \cite{FeldmannKS98}.

In \req{qsbasis} the $\sbs$ ($\qbq$) Fock component appears in the 
$\eta_q$ ($\eta_s$). These respective opposite Fock components  
lead to violations of the OZI rule if they were not suppressed. 
In order to achieve the mixing behaviour (\ref{qsbasis}),
(\ref{mixqs}) and, hence, strict validity of the OZI rule,
$\phi_{\rm opp}$ must be zero which implies
\be 
    \phi_8(x,\muF) = \phi_1(x,\muF) = \phi_q(x,\muF) = \phi_s(x,\muF)\,.
\label{da-rel}
\ee 
However, except the \da s assume the asymptotic form, this can only
hold approximately for a limited range of the factorization scale 
since the evolution of the \da s will generate differences between 
$\phi_1$ and $\phi_8$ and, hence, the respective opposite Fock 
components. In order to guarantee at least the approximate validity 
of the OZI rule and the quark-flavor mixing scheme as is required 
by phenomenology, we demand in our analysis of the transition form 
factor data that
\begin{equation}
\left|\frac{\phi_{\rm opp}(x,\muF)}{\phi_{AS}(x)} \right| \ll 1 \, ,
\label{eq:OZIreq}
\end{equation}
for any value of $x$.

\section{Determination of the distribution amplitudes}
\label{sec:form}
Before we turn to the analysis of the $P\gamma$ transition form
factor data \cite{CLEO97,L398} and the determination of the $\eta$ and
$\eta'$ \da s a few comments on the choice of the factorization and
renormalization scales are in order. A convenient choice of the
factorization scale \footnote{
A detailed discussion of the the 
role of the factorization scale and the resummation of corresponding 
logs is presented in Refs.\ \cite{MelicNP01,MelicNP01a}.}
is $\muF=Q^2$, it avoids the $\ln{\muF/Q^2}$ terms in 
\req{eq:THqqexp} and \req{eq:THggexp}. Another popular choice
is $\muF=Q^2/2$ which reflects the mean virtuality of the 
exchanged quark. This choice facilitates comparison with the pion
\da{} as determined in \cite{DiehlKV01} in exactly the same way
we are going to fix the $\eta$ and $\eta'$ \da s. For the
renormalization scale we choose $\muR=Q^2/2$ for which choice 
arguments have been given on the basis of a next-next-to-leading 
order calculation of the pion form factor \cite{MelicNP01}. 

The transition form factor is evaluated using  the two-loop expression 
for $\als$ with four flavors and
$\Lambda^{(4)}_{\overline{MS}}=305 \mev$ \cite{PDG98}. The numerical
values for the decay constants and mixing angles are given in 
\req{eq:num81}. As the starting scale of the evolution we take
$\muO= 1\gev^2$. 

\begin{figure}
\centerline{\includegraphics[height=7.5cm]{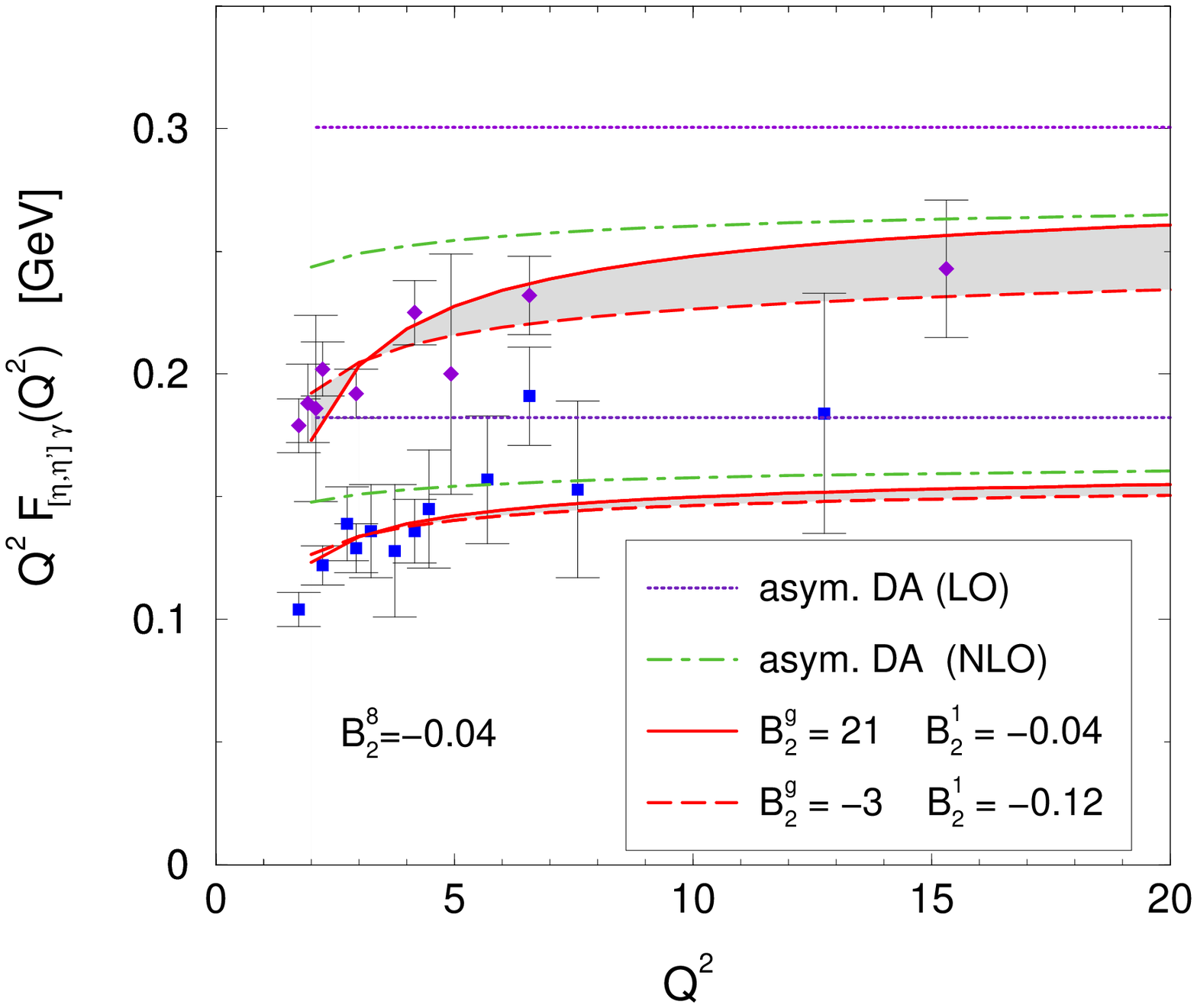}}
\caption{The scaled $P\gamma$ transition form factor vs.\ $Q^2$. 
Dotted (long-short dashed) lines represent the LO (NLO) predictions
for the asymptotic \da s. Solid (dashed) lines are results
obtained with $B_2^g(\muO)=21$ (-3), $B_2^1(\muO)=-0.04$ (-0.12)
and $B_2^8(\muO)=-0.04$ ($\muF=Q^2$, $\muR=Q^2/2$, $\muO=1\gev^2$).
The shaded areas indicate the range of the NLO predictions
for $B_2^1$ and $B_2^g$ inside the allowed region (see text). 
Data taken from \cite{CLEO97,L398} (rhombs represent the $Q^2 F_{\eta'
\gamma}$ data, squares the $Q^2 F_{\eta \gamma}$ ones).}
\label{f:etff}
\end{figure}

A comparison of the leading-twist NLO results evaluated from the 
asymptotic quark \da s (\ref{eq:phiAS}) (the gluon \da{} is zero in
this case) with experiment \cite{CLEO97,L398} is made in Fig.\ 
\ref{f:etff}. It reveals that the \da{}s cannot assume their
asymptotic forms for scales of the order of a few $\gev^2$; the 
prediction for the case of $\eta'$ lies about $10\%$ above the data. 
This parallels observations made for the case of the $\pi\gamma$ 
transitions \cite{DiehlKV01,KrollR96}. 

Next let us inspect the Gegenbauer expansion of the transition
form factor.
For $x$-independent factorization and renormalization scales the
integrations involved in \req{eq:Resetatff} and \req{eq:tffeta8} can be
performed analytically leading to the expansion 
\begin{eqnarray}
F^1_{P\gamma}(Q^{2})&=&  \frac{6 \, f_P^{1} \, C_{1}}{Q^2} \;
                 \bigg\{ 1 + B_2^1(\muF) + B_4^1(\muF) 
                         \nn             \\
              & &  - \frac53\frac{\als(\muR)}{\pi} \, \left[
             1 - B_2^1(\muF)\,
         \left(\frac{59}{72} -\frac{5}{6}\ln{\frac{Q^2}{\muF}}\right) 
       \right. \nn \\ & & \left. \: \:
              -B_4^1(\muF)\,
     \left(\frac{10487}{4500} -\frac{91}{75}\ln{\frac{Q^2}{\muF}}\right)  
                        \right. \nonumber \\ 
      & & \left.  \: \: + B_2^g(\muF)  
      \left(\frac{55}{1296}-\frac{1}{108}\ln \frac{Q^2}{\muF}\right)
       \right. \nn \\ & & \left. \: \:
     + B_4^g(\muF) 
      \left(\frac{581}{10125}-\frac{7}{675}\ln \frac{Q^2}{\muF}\right)
        \right] + \cdots \bigg\}  \, . \nonumber 
\nn \\ & &
\label{eq:Tffactors}
\end{eqnarray}
Particle independence of the \da s is used in this expansion. 
A similar expansion holds for the octet 
contribution with the obvious replacements $f_P^1 \to f_P^8$,  
$B_n^1\to B_n^8$, and $B_n^g \to 0$. 
The expansion of the octet contribution is analogous to
that one of the $\pi\gamma$ transition form factor
\cite{MelicNP01,DiehlKV01}.   

In the expansion \req{eq:Tffactors} one notes a strong linear 
correlation between $B_2^i$ and $B_4^i$, only the mild logarithmic 
$Q^2$ dependence due to evolution and the running of $\als$ 
restricts their values to a finite region in 
parameter space. The gluon contributions to the form factors are
strongly suppressed, they appear only to NLO and the numerical
factors multiplying their Gegenbauer coefficients are small. 
The coefficients $B_2^g$ and $B_4^g$ are also correlated.

\begin{figure}
\centerline{\includegraphics[height=5.8cm]{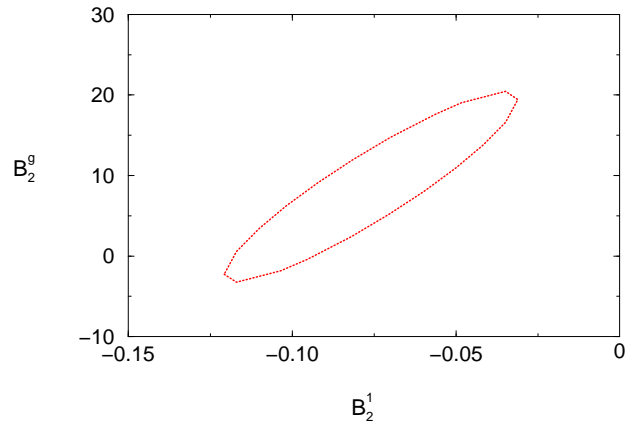}}
\caption{$1\sigma$ $\chi^2$-contour plot for the coefficients 
$B_2^1(\muO)$ and $B_2^g(\muO)$ obtained from a three-parameter fit to the 
CLEO and L3 data on the $\eta, \eta'$--$\gamma$ transition form
factors. Values of the Gegenbauer coefficients refer to 
$\muO=1 \gev^2$; the factorization scale is $\muF=Q^2$.}
\label{f:corr}
\end{figure}

\begin{figure}
\centerline{\includegraphics[height=6.7cm]{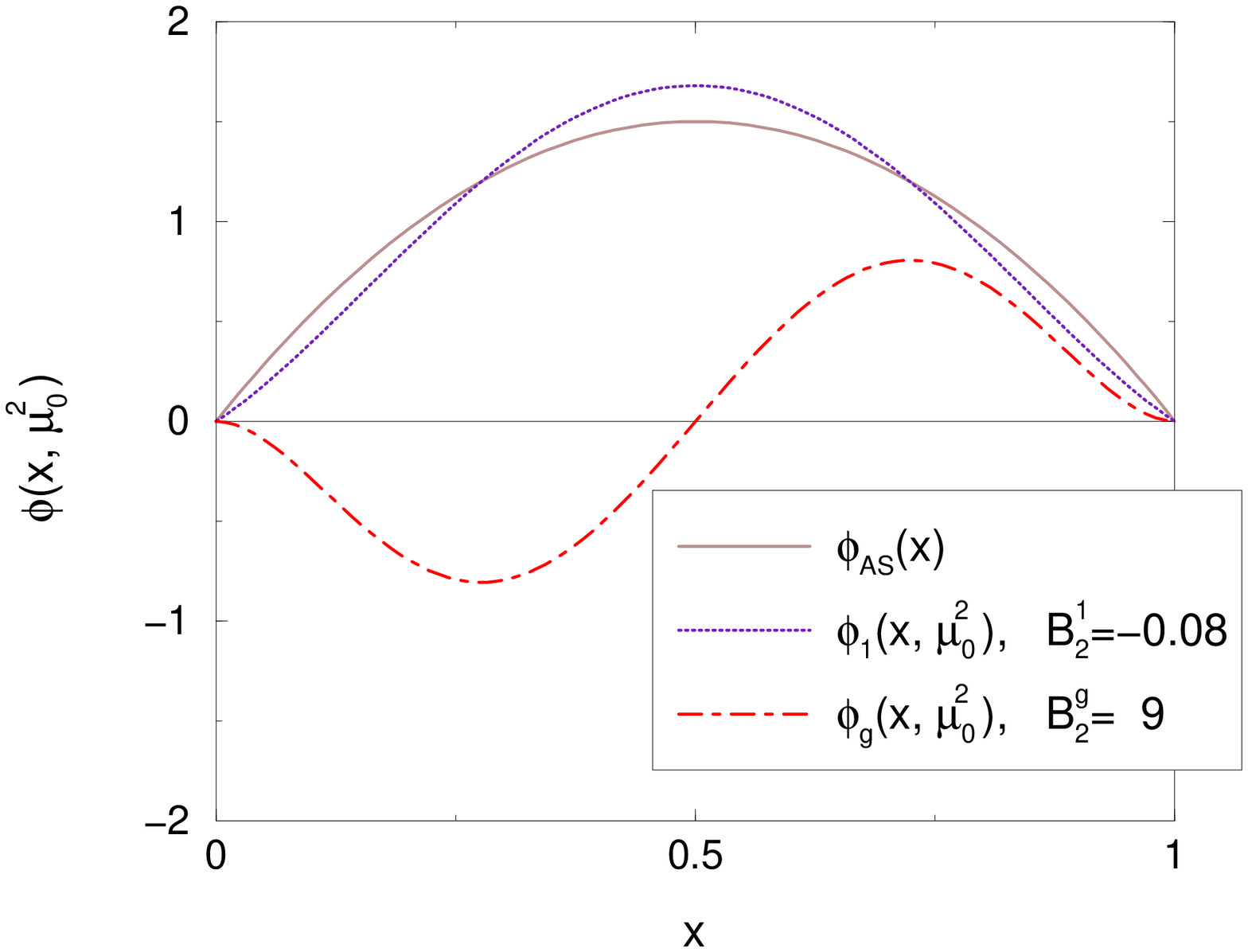}}
\caption{Flavour-singlet and gluon \da s at the scale $\muO=1\gev^2$
obtained using the face values $B_2^1$ and $B_2^g$
from \protect\req{eq:fit81g}. The asymptotic \da{} is included
for comparison.}
\label{f:DA0}
\end{figure}

With regard to these correlations and in view of the errors of the 
experimental data \cite{CLEO97,L398} as well as the
rather restricted range of momentum transfer in which they are
available, we are forced to truncate the Gegenbauer series at $n=2$. 
Truncating at $n=4$ does not lead to reliable results in contrast to 
the simpler case of the pion where this is possible \cite{DiehlKV01}. 
A fit to the CLEO and L3 data for  $Q^2$ larger then $2 \gev^2$ 
provides
\begin{eqnarray}
B_2^8(\muO) &=& -0.04 \pm 0.04 \, ,\nonumber \\[0.3cm]
B_2^1(\muO) & =& -0.08 \pm 0.04 \, , \nonumber \\[0.2cm]
B_2^g(\muO) & =& \phantom{-}9 \pm 12 
\, ,
\label{eq:fit81g}
\end{eqnarray}
where the values of the Gegenbauer coefficients are obtained 
for the factorization scale $\muF=Q^2$. 
We repeat that $\muO=1\gev^2$ and the gluonic Gegenbauer coefficient
is quoted for the normalization $\sigma=\sqrt{n_f/C_F}$.
For comparison we also determine the 
Gegenbauer coefficients for $\muF=Q^2/2$; the values found agree 
with those quoted in \req{eq:fit81g} almost perfectly. 
The quality of the fit is shown in Fig. \ref{f:etff}. 
The coefficients $B_2^1$ and $B_2^g$ are strongly correlated as can be 
seen from Fig.\ \ref{f:corr}. The results \req{eq:fit81g} satisfy 
$\sqrt{2}|B_2^8(\muF)-B_2^1(\muF)|/3 \ll 0.02$ for all
$\muF > \muO$. This meets the requirement \req{eq:OZIreq}, and, 
therefore no substantial violations of the OZI rule follow from our 
\da s. It moreover implies the approximative validity of the 
quark-flavor mixing scheme advocated for in Ref.\ \cite{FeldmannKS98}.
In Fig. \ref{f:DA0} we present the singlet and gluon
\da s at the scale $\muO$ obtained using the
face values from \req{eq:fit81g}. Both amplitudes
are end-point suppressed as compared to the asymptotic one.
This property holds for all values of
$B_2^1$ and $B_2^g$ inside the allowed region \req{eq:fit81g}.

The values of $B_2^1$ and $B_2^8$ agree with each other within errors 
as well as with the Gegenbauer coefficient $B_2^{\pi}(\muO)$
of the pion \da{} for which a value of $-0.06\pm0.03$ has been found
in \cite{DiehlKV01} from an analysis along the same lines as our one.
Thus, the three quark \da s are very similar. This result explains 
the observation made in \cite{Feldmann99} and mentioned by us at the end of 
Sec.\ II D that the data on three transition form factors fall on top 
of each other within errors if the form factors are scaled by their 
respective asymptotic results \req{f-asy}. The $\eta_c \gamma$ 
transition form factor, on the other hand, behaves differently \cite{L399}.
The $\eta_c$ mass provides a second large scale which cannot
be ignored in the analysis \cite{FeldmannK97a}. 

We emphasize that our results on the $\eta$ and $\eta'$  \da s are to 
be considered as estimates performed with the purpose of getting 
an idea about the magnitude of the gluon \da. As has been 
discussed in detail for the case of the $\pi\gamma$ transition 
form factor in \cite{DiehlKV01}, allowance of higher Gegenbauer
coefficients in the analysis will change the result on $B_2^\pi$,
essentially the sum of the $B_n^\pi$ is fixed by the data on
the transition form factor. This ambiguity also holds for the
case of the $\eta$ and $\eta'$. Taking a lower renormalization
scale than we do which may go along with a prescription for the
saturation of $\als$ and thus including effects beyond a 
leading-twist analysis, will also change the results for the 
Gegenbauer coefficients. Another source of theoretical uncertainties 
in our analysis is the neglect of power and/or higher-twist 
corrections. Thus, for instance, in Refs.\ \cite{FeldmannK97,JKR} 
the LO modified perturbative approach \cite{Sterman} has been applied 
where quark transverse degrees of freedom and Sudakov
suppressions are taken into account. In this case the asymptotic
\da s lead to good agreement with the data on the 
transition form factors.
\section{Comments on other hard reactions} 
\label{sec:appl}
In this section we make use of the results obtained in the 
preceding sections and calculate other hard processes involving 
$\eta$ and $\eta'$ mesons in order to examine the role of the 
$gg$ Fock component further.

\subsection{Electroproduction of $\eta, \eta'$ mesons} 
\label{sec:DVEM}
\begin{figure}
\centerline{\includegraphics[height=3.5cm]{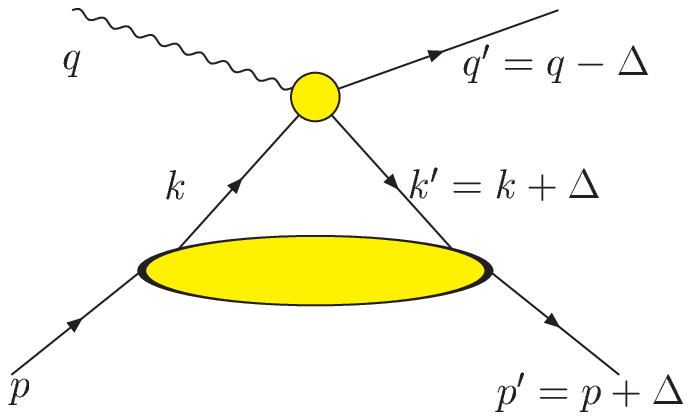}}
\caption{The handbag-type diagram for meson electroproduction off protons.
The large blob represents a generalized parton distribution, the small 
one the subprocess $\gamma^*_L q \rightarrow P \, q$. The momentum
transfer is $t=\Delta^2$.}
\label{f:handbag}
\end{figure}
As a first application of the gluon \da{} extracted from the
$\eta\gamma$ and $\eta'\gamma$ transition form factors we calculate
deeply virtual electroproduction of $\eta$ and $\eta'$ mesons
off protons. It has been shown \cite{Radyushkin96,CollinsFS96} that 
for large virtualities of the exchanged photon, $Q^2$, and small 
momentum transfer from the initial to the final proton, $t$, 
electroproduction of pseudoscalar mesons is dominated by
longitudinally polarized virtual photons and the process amplitude 
factorizes into a parton-level subprocess $\gamma^*_L\, q \to P q$ 
and soft proton matrix elements which represent generalized parton 
distributions \cite{MullerRGDH94}, see Fig.\ \ref{f:handbag}.  
The meson is generated by a leading-twist mechanism, i.e., by the 
transition $\qbq \to P$ mediated through the exchange of a hard 
gluon. For the production of $\eta$ and $\eta'$ mesons, however, one 
has to consider the gluon Fock component as well which, in contrast 
to the case of the transition form factors, contributes to the same 
order of $\als$ as the  $\qbq_i$ components. The gluonic 
contribution has not been considered in previous calculations of 
the electroproduction cross sections \cite{MankiewiczPW97,HuangK00}.

\begin{figure}[t]
\centerline{\includegraphics[width=8cm]{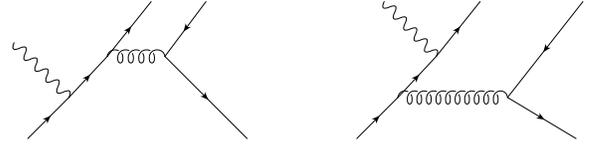}}
\caption{Sample leading order Feynman diagrams that contribute to the 
subprocess amplitude
$\gamma^*_L q \rightarrow \qbq_i \, q$.}
\label{f:DVEMqq}
\end{figure}

The helicity amplitude for the process $\gamma^*_L\, p \to P p$
is again decomposed into flavor octet and singlet components,
$\qbq_i \to P$
\ba
{\cal M}^{Pi}_{0\pm,0\pm}&=& 
                \sum_a e e_a C_a^{\,i} \sqrt{1-\xi^2} \int_{-1}^1\, 
                 \frac{d\bar{x}}{\sqrt{\bar{x}^2 -\xi^2}} \,                 
                   {\cal H}^{Pi}_{0\pm,0\pm} \nn\\
             &&\times  \left[\widetilde{H}^a(\bar{x},\xi,t) 
           - \frac{\xi^2}{1-\xi^2}\widetilde{E}^a(\bar{x},\xi,t)\right]\,,
\label{eq:qeamp}
\ea
where $\widetilde{H}^a$ and $\widetilde{E}^a$ are the axial vector and
pseudoscalar generalized parton distributions for the emission and 
reabsorption of quarks of flavor $a$. The $C_a^i$ are flavor factors 
for the $\qbq_i$ components of the meson $P$; they can be read off from
\req{eq:ffactors}. 
The quark subprocess amplitudes $\gamma_L^* q \to \qbq_i\, q$ are
calculated from the LO Feynman diagrams for which examples are shown 
in Fig. \ref{f:DVEMqq} \cite{HuangK00}
\ba
{\cal H}^{Pi}_{0\pm,0\pm}(\hat{s},{t},Q^2) &=& \pm\, 4\pi \als(\muR) 
                           \frac{C_F}{N_c}\, f_P^i\,
                 \frac{Q\sqrt{-\hat{u}\hat{s}}}{Q^2+\hat{s}} 
\nn\\ && \times
\int_0^1\,d\tau\, \frac{\phi_i(\tau,\muF)}{(1-\tau)Q^2-\tau t}\,
\nn\\ && \times
              \left[ 1 -\frac{\hat{u}}{\hat{s}} 
                    + \frac1{1-\tau}\frac{t}{\hat{u}} \right]\,.            
\ea
They are expressed in terms of the subprocess Mandelstam variables 
$\hat{s}, \hat{u}, \hat{t}=t$ where $\hat{s}+t+\hat{u}=-Q^2$, and hold 
for any value of $Q^2$ and $t$. For the deeply virtual kinematical 
region of large $Q^2$ and $-t \ll Q^2$, it is more appropriate to use 
the scaling variables $\xi$ and $\bar{x}$. The skewness is defined by
the ratio of light-cone plus components of the incoming ($p$) and 
outgoing ($p'$) proton momenta
\begin{equation}
\label{eq:skewedness}
\xi= \frac{(p-p^{\,\prime})^+}{(p+p^{\,\prime})^+}\,.
\end{equation}
For large $Q^2$ the skewness is related to  $x$-Bjorken by
$\xi\simeq x_{Bj}/2$. The average momentum fraction the
emitted and reabsorbed partons carry, is defined as 
\be
\bar{x}= \frac{(k+k')^{\,+}}{(p+p')^{\,+}}\,.
\ee
Here, $k$ and $k'$ are the momenta of the emitted and reabsorbed
partons, respectively. For $-t \ll Q^2$ the Mandelstam
variables are related to the skewness and the average momentum
fraction 
\newpage
\begin{widetext}
\begin{figure*}
\centerline{\includegraphics[width=12cm]{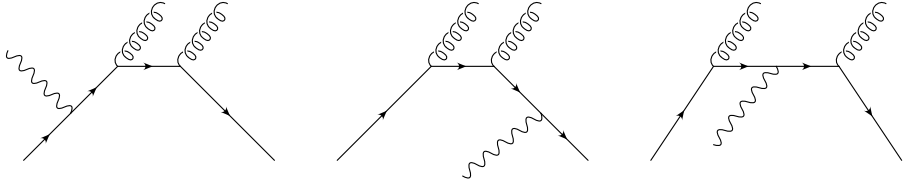}}
\caption{Representative LO Feynman diagrams that contribute to the 
subprocess amplitude
$\gamma^*_L q \rightarrow gg \, q$.}
\label{f:DVEMgg}
\end{figure*}
\end{widetext}
\be
\hat{s}=\frac{Q^2}{2\xi}(\bar{x}-\xi)\,, \qquad  
\hat{u}=-\frac{Q^2}{2\xi}(\bar{x}+\xi)\
\, .
\ee 
Rewriting the subprocess amplitude in terms of $\xi$ and $\bar{x}$
and inserting the result into the factorization formula \req{eq:qeamp},
one arrives at the well-known result for the leading-twist
contribution to deeply virtual electroproduction of pseudoscalar mesons 
\cite{MankiewiczPW97}
\ba
\lefteqn{{\cal M}^{Pi}_{0\pm,0\pm}(Q^2,\xi,t \simeq 0)} 
\nn \\ & = & \pm \frac{4\pi\als(\muR)}{Q}\,
                            \frac{C_F}{N_c}\, f_P^i \sqrt{1-\xi^2}
                      \int_0^1 d\tau \frac{\phi_i(\tau,\muF)}{\tau}
              \,   
\nn \\ & & \times \sum_a e e_a C_a^i \: 
       \int_{-1}^1\, {d\bar{x}}
	  \left[ \frac{1}{\bar{x}+\xi-i\veps} +
	    \frac{1}{\bar{x}-\xi+i\veps} \right]
\nn \\ & & \times  
             \left[ \widetilde{H}^a(\bar{x},\xi,t) -
	  \frac{\xi^2}{1-\xi^2}\widetilde{E}^a(\bar{x},\xi,t)\right]\,.
\label{eq:eamp}
\ea

Next we calculate the subprocess amplitude for the gluonic component
of the meson, $\gamma^*_L\, q \to gg \, q$. 
There are six graphs that contribute to the subprocess. Three
representative ones are depicted in Fig.\ \ref{f:DVEMgg}, the other
three ones are obtained from these by interchanging the gluons. 
We find for that subprocess amplitude the result
\ba
{\cal H}^{Pg}_{0\pm,0\pm}(\hat{s},t,Q^2)&=& 
            \mp 4\pi \als(\muR)\, \frac{f_P^1}{\sqrt{n_f}}\, 
                   \frac{C_F}{N_c}\,
            \frac{Q}{Q^2+\hat{s}}\, \frac{-t}{\sqrt{-\hat{u}\hat{s}}}   
\nn \\ & & \times
                            \,\int_0^1 d\tau \,
			    \frac{\phi_g(\tau,\muF)}{\tau^2(1-\tau)} \,.
\label{eq:gamp}
\ea
In deriving this expression we made use of the antisymmetry of the
gluon \da{} \req{eq:DAsim}. The gluonic contribution to the 
$\gamma^*_L\, p\to P p$ helicity amplitudes reads 
\ba
{\cal M}^{Pg}_{0\pm,0 \pm} 
              &=& \sum_a e e_a \sqrt{1-\xi^2}\,  
          \int_{-1}^1 \, \frac{d\bar{x}}{\sqrt{\bar{x}^2-\xi^2}} \, 
              \,{\cal H}^{Pg}_{0\pm,0\pm} \nn\\[0.2cm]
             & \times& \left[\widetilde{H}^a(\bar{x},\xi,t) 
       - \frac{\xi^2}{1-\xi^2}\,\widetilde{E}^a (\bar{x},\xi,t)\right]
\, .
\label{eq:Mpg}
\ea 
The full  $\gamma^*_L\, p\to P p$ amplitudes are the sum of the
flavor octet and singlet contributions \req{eq:eamp} and the gluonic
one \req{eq:Mpg}. In the deeply virtual region, however,
the gluon contribution is suppressed by $t/Q^2$ as one readily
observes from \req{eq:gamp}. It is, therefore, to be considered as a
power correction to the leading quark contribution \req{eq:eamp} and
is to be neglected in a leading-twist analysis of deeply virtual
electroproduction of $\eta$ and $\eta'$ mesons. 

One may also consider wide-angle photo- and electroproduction of
$\eta$ and $\eta'$  mesons. Using the methods proposed in
\cite{DiehlFJK98} for wide-angle Compton scattering, one can show that 
for wide-angle photo- and electroproduction of pseudoscalar mesons the
factorization formulas \req{eq:qeamp} and \req{eq:Mpg} hold as well
provided $-t$ and $-u$ are large as compared to the square of the
proton mass and $Q^2\ll -t $ \cite{HuangK00}. To show that one has to
work in a symmetric frame in which the skewness is zero. One can also
show that, in this situation, $\hat{s}$ and $\hat{u}$ are approximate 
equal to the Mandelstam variables for the full process, $s$ and $u$, 
respectively. Thus, in the wide-angle region and for $Q^2\ll -t, s$
but non-zero, \req{eq:qeamp} and \req{eq:Mpg} simplify to
\ba
{\cal M}^{Pi}_{0\pm,0\pm}(s, t, Q^2 \ll -t) &=& 
                   e\, {\cal H}^{Pi}_{0\pm,0\pm} \sum_a e_a
                                    C_a^i R_A^a(t)\,, \nn\\
{\cal M}^{Pg}_{0\pm,0\pm}(s, t, Q^2 \ll -t) &=& 
                       e\, {\cal H}^{Pg}_{0\pm,0\pm} \sum_a e_a
                                        R_A^a(t)\,, 
\nn \\
\ea
where the form factors $R_A^a$ represent $1/\bar{x}$ moments of the
generalized parton distributions $\widetilde{H}^a$ at zero
skewness. These form factors also contribute to wide-angle Compton
scattering \cite{DiehlFJK98}.  
The amplitudes for transversally polarized photons can be obtained
analogously. In contrast to the case of deeply virtual electroproduction
\cite{MankiewiczP99},
factorization for these amplitudes holds in the wide-angle 
region, too.

\begin{figure}
\centerline{\includegraphics[height=5.5cm]{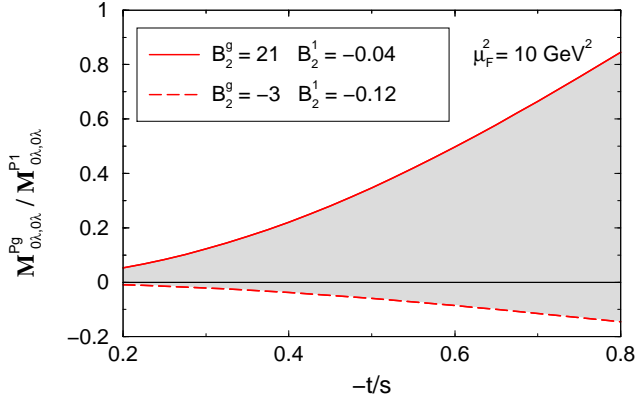}}
\caption{Ratio of gluon and flavor-singlet quark amplitudes for
wide-angle electroproduction of $\eta$ or $\eta'$ mesons
($\muF=10\gev^2$). The shaded area indicates the range of predictions
evaluated from $B_2^1(\muO)$ and $B_2^g(\muO)$ inside the allowed 
region according to Fig. \protect\ref{f:corr}.}
\label{f:ratio}
\end{figure}

In order to estimate the size of the gluon contribution to
wide-angle electroproduction of $\eta, \eta'$ mesons, we plot 
in Fig.\ \ref{f:ratio} the ratio 
\ba
\frac{{\cal M}^{Pg}_{0\pm,0\pm}} {{\cal M}^{P1}_{0\pm,0\pm}}&=&
              \frac{- t^2} {2s^2 +t^2 + ts} 
\nn \\ & & \times
             \int_0^1 d\tau \frac{\phi_g(\tau,\muF)} {\tau^2(1-\tau)}
	     \left[\int_0^1 d\tau \frac{\phi_1(\tau,\muF)} {\tau} \right]^{-1}\,,
\nn \\ & &
\ea
evaluated from the \da s \req{eq:fit81g}
for which the ratio of the integrals is $\simeq -5\, B_2^g(\muF)/18$.
The ratio may be large in particular in the backward hemisphere. 
Thus, at least for electroproduction of $\eta'$ mesons the $gg$ Fock
component should be taken into account for sufficiently large momentum
transfer. For the production of the $\eta$ meson it plays a minor 
role since $\eta$ production is dominated by the flavor-octet 
contribution ($f_\eta^1/f_\eta^8=0.16$). 
Note, however, that the normalization of the
meson electroproduction in both the regions, 
the deeply virtual and the wide-angle one, is not well understood 
in the kinematical region accessible to present day experiments. 

\subsection{The $g^* g^* P$ vertex}
\label{sec:etagg}
A reliable determination of the $g^* g^* \eta'$ vertex is of
importance for the calculation of a number of decay processes such as 
$B \to \eta' K$, $B \to \eta' X_s$, or of the hadronic production
process $p p  \to \eta' X$. The $g^* g^* \eta'$ vertex has been
calculated by two groups recently \cite{MutaY99,AliP00}. We 
reanalyze this vertex to leading-twist order using our set of
conventions. This will allow us to examine the previous calculations,
and provide predictions for $Pg^*$ transition form factor using
the Gegenbauer coefficients \req{eq:fit81g} in the \da s. 

We define the gluonic vertex in analogy to the electromagnetic one ,
see \req{eq:Gmu}, as
\begin{equation}
    \Gamma^{\mu \nu}_{a b}=
  i  \; F_{Pg^*}(\ov{Q}^{\,2},\omega) 
    \; \delta_{ab} \; \eps^{\mu \nu \alpha \beta} 
    \; q_{1\alpha} q_{2\beta}
\label{eq:Gmunuab}
\end{equation}
where $q_1$ and $q_2$ denote the momenta of the gluons now and 
$a$ and $b$ label the color of the gluon. It is
evident that the transition to a colorless meson requires the same
color of both the gluons.  We consider space-like gluon virtualities
for simplicity; the generalization to the case of time-like gluons is
straightforward. We introduce an average virtuality and an asymmetry
parameter by
\begin{equation}
\ov{Q}^{\,2}=-\frac{1}{2} (q_1^2+q_2^2)
\, ,
\qquad
\omega=\frac{q_1^2-q_2^2}{q_1^2+q_2^2} \, .
\label{eq:Qomega}
\end{equation}
The values of $\omega$ range from $-1$ to $1$,
but due to Bose symmetry the transition form factor is
symmetric in this variable:
$F_{Pg^*}(\ov{Q}^{\,2},\omega)= F_{Pg^*}(\ov{Q}^{\,2},-\omega)$.

\begin{figure}
\centerline{\includegraphics[height=6.5cm]{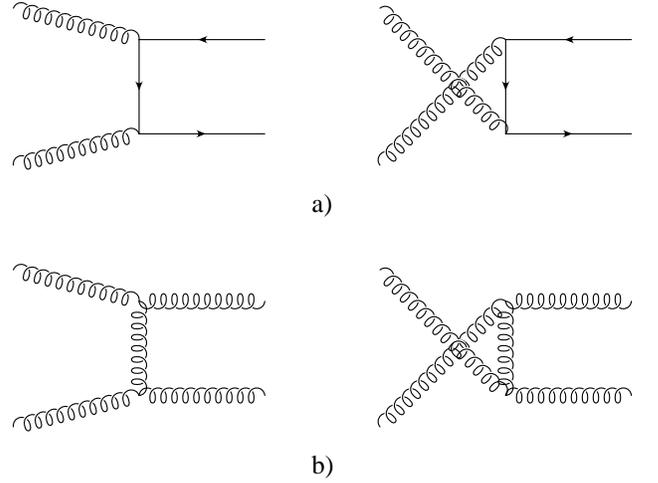}}
\caption{ Relevant lowest order Feynman diagrams for the $g^*g^*\to \qbq$ (a)
and  $g^*g^*\to gg$ subprocess (b).}
\label{f:qqgggggg}
\end{figure}

The calculation of the transition form factor to leading twist
accuracy and lowest order in $\alpha_s$ parallels that of the meson-photon
transition form factor which we presented in some detail in Sec.\ 
\ref{sec:singlet}. In contrast to the electromagnetic case, however,
already to the lowest order in $\alpha_s$ the two partonic
subprocesses $g^* g^* \to q \bar{q}$ and $g^* g^* \to gg$ contribute. 
The relevant Feynman diagrams are shown in Fig. \ref{f:qqgggggg}.
There are a few more diagrams which involve the triple and quadruple
gluon vertices. The contributions from these diagrams are separately zero
when contracted with either the $\qbq$ or the $gg$ projectors
\req{eq:qq}, \req{eq:ggNEW}.
The following result for the $Pg^*$ transition form factor can
readily be obtained 
\begin{eqnarray}
F_{Pg^*}(\ov{Q}^{\,2},\omega)
 &=& 4 \pi \als(\muR) \; \frac{f_{P}^1}{\ov{Q}^{\,2}} \;
            \: \frac{\sqrt{n_f}}{N_c} \; 
\nn \\ & & \times
            \left[ A_{q\overline{q}}(\omega) + \frac{N_c}{2 n_f} \,
                    A_{gg}(\omega) \right] +{\cal O}(\alpha_s^2)\, ,
\nn \\[0.15cm] &&
\label{eq:ourFggeta}
\end{eqnarray}
where 
\begin{eqnarray}
 A_{q\overline{q}}(\omega) &=& \int_0^1 dx \, \phi_1(x,\mu_F^2)
                      \; \frac{1}{1-\omega^2 (1-2 x)^2} \, ,
                                                            \nn\\
 A_{gg}(\omega) &=& \int_0^1 dx \, \frac{\phi_g(x,\mu_F^2)}
           {x \overline{x}} \; \frac{1-2 x}{1-\omega^2 (1-2 x)^2} \, .
\nn \\ & &
\label{eq:ourAqqgg}
\end{eqnarray}
There is no contribution from the $\qbq_8$ component to
this vertex.

Inserting the Gegenbauer expansions \req{eq:solphi} into
\req{eq:ourAqqgg} the integrals can be performed
analytically term by term analogously to \req{eq:Tffactors} resulting
in the expansions 
\begin{eqnarray}
 A_{q\overline{q}}(\omega) &=& 
   c_0(\omega) + c_2(\omega) \; B^{1}_2(\muF) + \cdots
\, ,
 \nonumber \\
 A_{gg}(\omega) &=& g_2(\omega) \; B_2^{g}(\muF) + \cdots
\, ,
\label{eq:omA} 
\end{eqnarray}
where
\begin{eqnarray}
 c_0(\omega) &=& \frac{3}{2\omega^2}
       \left[ 1 
  - \frac{1}{2 \omega} \left(1 - \omega^2 \right)
               \ln \frac{1+\omega}{1-\omega} \right]
\, ,
 \nonumber \\
 c_2(\omega) &=& \frac{3}{4\omega^4}
       \left[ 15 -13 \omega^2 
  - \frac{3}{2 \omega} \left( 5 - 6 \omega^2 + \omega^4 \right)
               \ln \frac{1+\omega}{1-\omega} \right]
\, ,
 \nonumber \\
 g_2(\omega) &=& \frac{-5}{12\omega^4}
       \left[ 3 - 2 \omega^2 
  - \frac{3}{2 \omega} \left(1 - \omega^2 \right)
               \ln \frac{1+\omega}{1-\omega} \right]
\, .
\nn \\ & &
\label{eq:c0c2g2} 
\end{eqnarray}
\begin{figure} 
\begin{center} 
\includegraphics*[width=7cm]{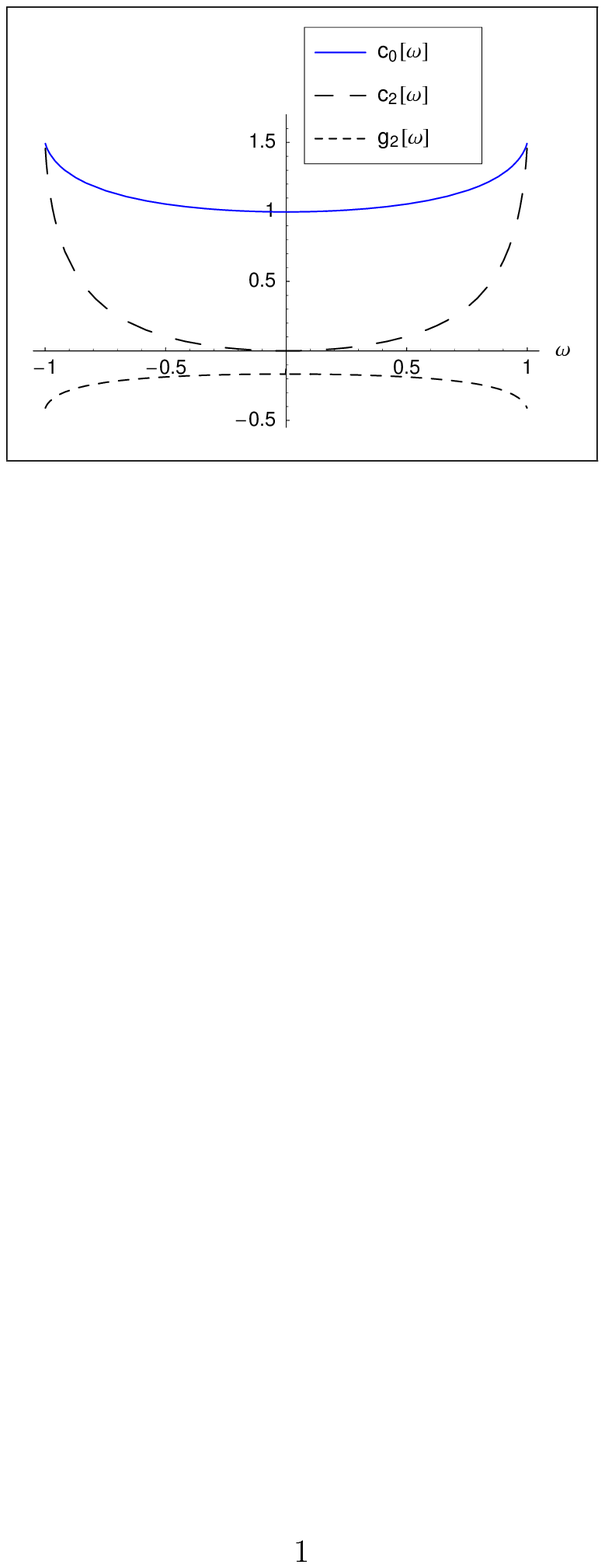} 
\end{center} 
\caption{Functions $c_0$, $c_2$, and $g_2$, defined in Eq.\ 
\protect\req{eq:c0c2g2}, vs.\ $\omega$.}
\label{f:c0c2g2}
\end{figure} 
The behaviour of functions $c_0(\omega)$, $c_2(\omega)$, and
$g_2(\omega)$
is illustrated in Fig. \ref{f:c0c2g2}. Examining the function
$c_2(\omega)$ and Eq.\ \req{eq:omA}, one notice that the form factors 
become increasingly less sensitive to the coefficients $B_2^1(\muF)$
with decreasing $|\omega|$.
This behaviour is characteristic of all functions
$c_n(\omega)$ ($n >0$) \cite{DiehlKV01}. On the other hand, the 
functions $c_0(\omega)$ and $g_2(\omega)$ 
do not depend so drastically on $\omega$ and 
they are non-zero at $\omega=0$. One can easily show that all
$g_n(\omega)$, for $n>0$ and even, possess this property.

Let us discuss two interesting limiting cases.
For $\omega \ll 1$, i.e., for $q_1^2 \approx q_2^2$,
the form factors behave as 
\begin{eqnarray}
F_{Pg^*}(\ov{Q}^{\,2},\omega)&=& \frac{4 \pi \als(\muR)}{\sqrt{3}\; \ov{Q}^{\,2}} \; 
                          f_{P}^1 \; \left[1- \frac{1}{12} B_2^g(\muF)
\right. \nn \\ & & \left.
                    + \frac{1}{5} \omega^2 \left(1+\frac{12}{7} B_2^1(\muF) 
                       -\frac{5}{28}B_2^g(\muF)\right) \right] \nn\\[3mm]
                          && +\, {\cal O}(\omega^4,\alpha_s^2)\, .
\label{eq:Fsmallom}
\end{eqnarray}
Thus, the limiting value for $\omega \to 0$ is sensitive to the form
of the gluon distribution amplitude while it does not depend on the
Gegenbauer coefficients of the quark one. This is to be contrasted
with the $P\gamma^*$ transition form factor which, according to 
\cite{DiehlKV01}, is independent of both the quark and the gluonic 
Gegenbauer coefficients in the limit $\omega \to 0$.

\begin{figure}
\begin{center} 
\begin{tabular}{c}
\includegraphics[height=5.5cm]{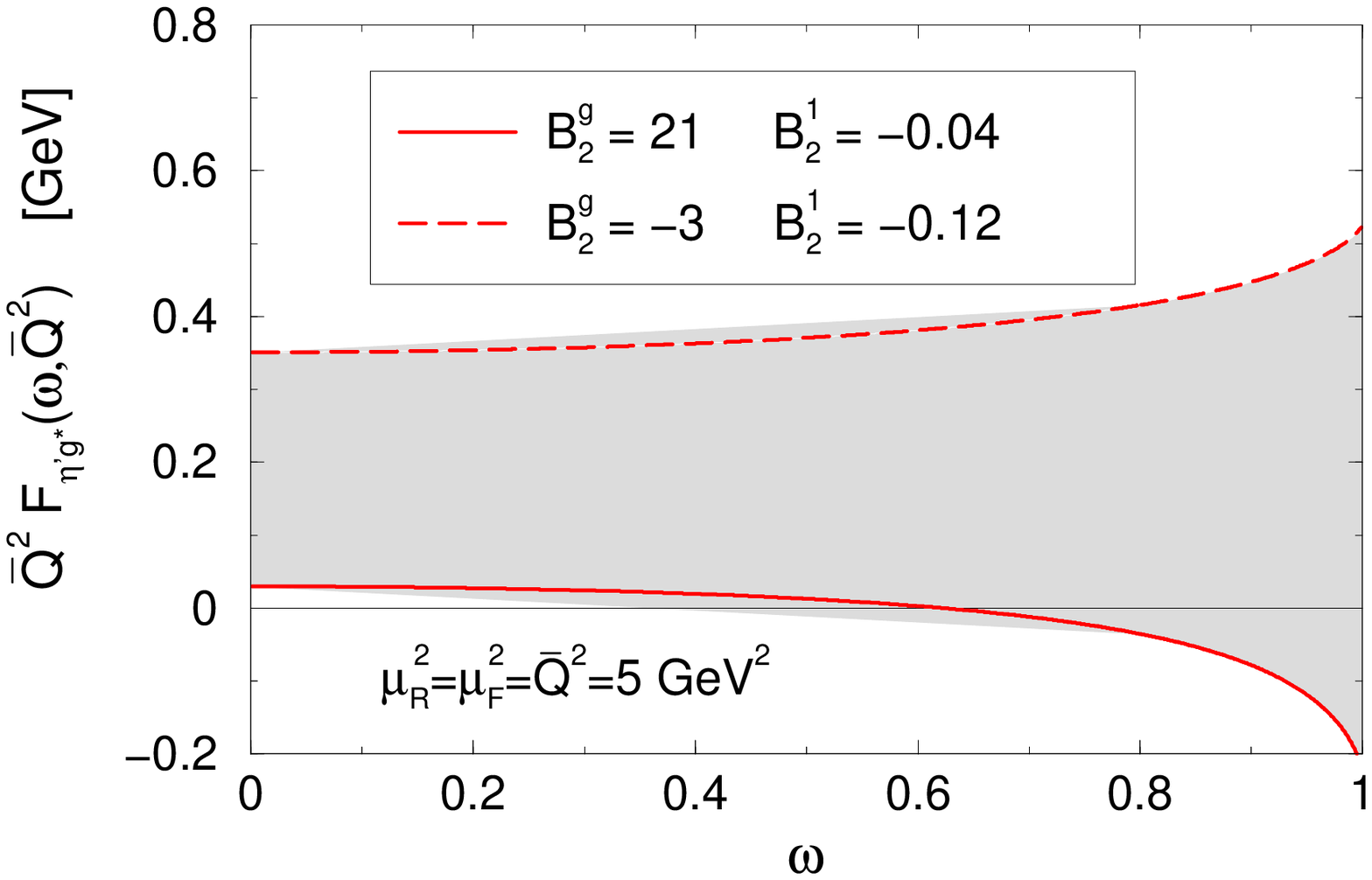} \\
\includegraphics[height=5.5cm]{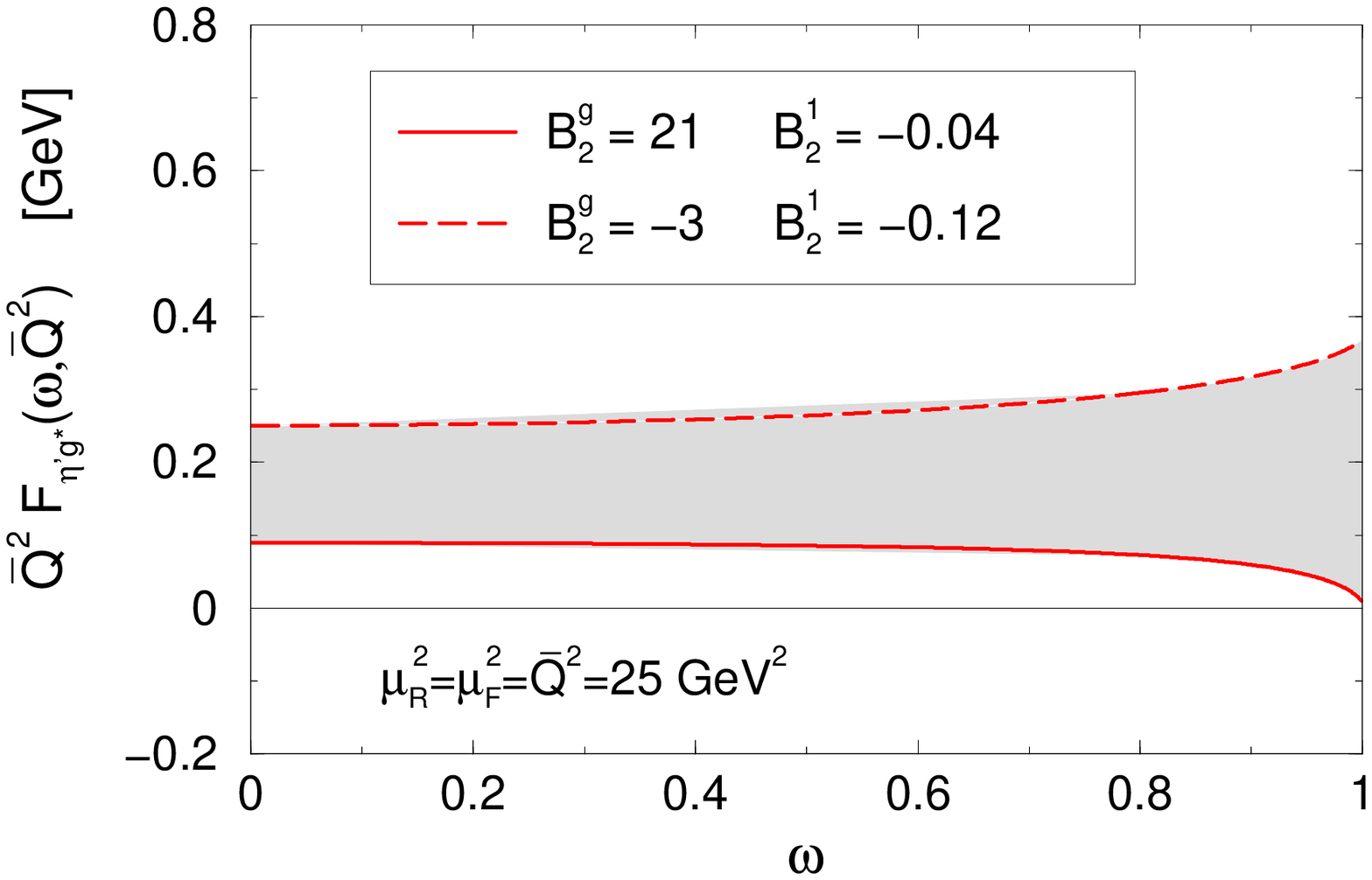}
\end{tabular}
\end{center}
\caption{Predictions for the $\eta'g^*$ transition form factor
as a function of $\omega$ for two values of $\ov{Q}^2$.
The shaded areas indicate the range of predictions
evaluated from $B_2^1(\muO)$ and $B_2^g(\muO)$ 
inside the allowed region according to Fig. \protect\ref{f:corr}.}
\label{f:ggformf}
\end{figure}

For $\omega \to \pm 1$, i.e., in the limit where one of the gluons 
goes on-shell, the $Pg$ transition form factor becomes
\begin{eqnarray}
F_{Pg}(Q^2,\omega=\pm 1)&=& \frac{4 \sqrt{3} \pi \als(\muR))}{Q^2} \; 
               f_{P}^1 \; 
 \nn \\ & & \times
\left[1+B_2^1(\muF)- \frac{5}{36} B_2^g(\muF)
               \right] 
 \nn \\ & & 
+\, {\cal O}(\alpha_s^2)\, ,
\label{eq:Fom1}
\end{eqnarray}
where $Q^2=-q_1^2\, (-q_2^2)$ as in the electromagnetic case.
In Fig.\ \ref{f:ggformf} we display our predictions for the scaled
$\eta'g^*$ transition form factor evaluated from the \da s determined
in Sec.\ \ref{sec:form}, choosing $\muF=\muR=\ov{Q}^2$. 
Given the large difference in the magnitude of $B_2^1(\muO)$ and
$B_2^g(\muO)$, see \req{eq:fit81g}, we observe a strong sensitivity of the
$Pg$ transition form factors on the gluon \da{} in contrast to the
electromagnetic case. Due to the badly determined coefficient $B_2^g$
the uncertainties in the predictions for $F_{\eta'g^*}$ are large.  
Because of the smallness of the mixing angle $\theta_1$,
see \req{mix81} and \req{eq:num81}, the $\eta g^*$ transition form
factor is much smaller then the $\eta'g^*$ one. The ratio of the two form
factors, $F_{\eta g^* }(\ov{Q}^2,\omega)/ F_{\eta'g^*}(\ov{Q}^2,\omega)$
is given by $-\tan \theta_1$. This result offers a way to measure the
angle $\theta_1$ as has been pointed out in \cite{FeldmannKS98}.

Let us compare our results for the $\eta'g^*$ transition form factors
with those presented in Refs.\ \cite{MutaY99,AliP00}. First we remark
that there is perfect agreement for the contribution from the meson's
$\qbq_1$ component. As for the contribution from the gluonic component
we differ by a factor $1/(2n_f)$ from Refs.\ \cite{MutaY99,AliP00}%
\footnote{We corrected a typo in \cite{MutaY99} where only the case of
$\omega=1$ has been dealt with - the relative sign
between the contributions from the two Feynman diagrams shown in Fig.\
\ref{f:qqgggggg} (b) should be minus. Moreover, in this work Ohrndorf's results
\cite{Ohrndorf81} for the anomalous dimensions are used which are
flawed while they have the same normalization as in
\req{eq:andimNEW}.}. Furthermore, in \cite{AliP00}, there is an
additional factor of $\omega$ multiplying the gluonic term rendering
it antisymmetric in $\omega$ in conflict with Bose symmetry.
We suspect 
that a gluonic projector $\sim \veps^{\mu\nu\alpha\beta} q_{1\alpha}
q_{2\beta}/\ov{Q}^2$ is used in \cite{AliP00} which turns into $\sim
\omega \veps^{\mu\nu}_\perp$ in a frame where the meson moves along
the 3-direction. This is in conflict with
(\ref{eq:helzero}), (\ref{eq:epsmunu}) except at $\omega=1$.

The origin of the missing factor $1/(2n_f)$ is not easy to
discover since in Refs.\ \cite{MutaY99,AliP00}
the form of the gluonic projector is not specified. 
Given the anomalous dimensions quoted in 
\cite{MutaY99,AliP00}, which are the same as in
\req{eq:andimNEW}, this incriminated factor
cannot be assigned to a particular normalization of the gluonic
projector, \req{eq:ggNEW} must be applied. 
On the other hand, using $\sigma=1/(2\sqrt{n_f\,C_F})$ 
as the normalization of the gluonic projector, 
the results for the transition form factors given
in \cite{MutaY99,AliP00} would be correct 
(ignoring the problem with the factor $\omega$ in \cite{AliP00}), 
provided the corresponding anomalous dimensions are applied, 
see \req{eq:gammasigma}, 
and they differ from the ones quoted in these papers.
Hence, the quoted anomalous dimensions and the result
for the gluon part of the hard-scattering amplitude
seem not to be in agreement.

In Ref.\ \cite{DiehlKV01} the leading term of the expansion
\req{eq:Fsmallom} has been derived from the results presented in
\cite{AliP00} and it therefore disagrees with our result. 
\section{Summary}
\label{sec:summ}
In this work we have investigated the two-gluon Fock components of the
$\eta$ and $\eta'$ mesons to leading-twist accuracy. Since the
integral over the gluon \da{} is zero, see \req{gda-nrom}, there is no
natural normalization of it in contrast to the case of the $\qbq$ \da s. 
Any choice of this normalization goes along with corresponding
normalizations of the anomalous dimensions and the projector of a
two-gluon state onto a pseudoscalar meson. We have set up a
consistent set of conventions for the three quantities which is
imperative for leading-twist calculations of hard exclusive reactions
involving $\eta$ and/or $\eta'$ mesons. We have also compared 
this set with
other conventions to be found in the literature.

As an application of the two-gluon components we have calculated the
flavor-singlet part of the $\eta\gamma$ and $\eta'\gamma$ transition
form factors to NLO in $\als$ and explicitly shown the cancellation 
of the collinear singularities present in the hard scattering
amplitude with the UV one occurring in the unrenormalized \da s. 
Assuming particle independence of the
\da s, we have employed the results for the transition form factors in
an analysis of the available data \cite{CLEO97,L398} and determined 
the Gegenbauer coefficients to order $n=2$ for the three remaining  
\da s, the flavor octet, singlet and gluon one. 
The numerical results for the \da s 
quoted for $\sigma=\sqrt{n_f/C_F}$ 
are in agreement
with the quark flavor mixing scheme proposed in \cite{FeldmannKS98}.

The value for the lowest order gluonic Gegenbauer coefficient is subject to a
rather large error since the contributions from the two-gluon Fock
components
to the transition form factors are suppressed by $\als$ as compared to
the $\qbq$ contributions. This suppression does not necessarily occur 
in other hard exclusive reactions; examples of such reactions,
discussed by us briefly, are deeply
virtual and wide-angle electroproduction of $\eta$ or $\eta'$ mesons
as well as the $g^* g^* \eta (\eta')$ vertex. The latter two reactions, 
as it has turned out, are actually quite sensitive to the two-gluon
components and future data for them should allow to pin down the gluon
\da{} more precisely than it is possible from the transition form factor
data. Other hard exclusive reactions which may be of relevance to our
considerations are, for instance, the decays $\chi_{cJ}\to \eta\eta,
\eta'\eta'$ \cite{BolzKS96,BaierG85a} or $B\to \eta^{(}{}'{}^{)} K^{(*)}$
\cite{beneke}. Last not least we would like to mention that the 
two-gluon components of other flavor-neutral mesons or even those of
glueballs \cite{carlson} can be studied in full analogy to the 
$\eta$-$\eta'$ case. 

\acknowledgments

We wish to acknowledge discussions with M.\ Beneke, M.\ Diehl, T.\
Feldmann, D.\ M\"{uller} and A. Parkhomenko.
This work was supported by Deutsche Forschungs Gemeinschaft
and partially supported by the Ministry of Science and Technology
of the Republic of Croatia under Contract No. 0098002.

\begin{appendix}
\section{Definitions of meson states and \da s}
\label{app:convdef}
The flavor content of the neutral pseudoscalar meson states  
we are interested in, is taken into account by
\begin{equation}
\begin{array}{llcl}
\pi^0 : & \displaystyle
\frac{1}{\sqrt{2}} (u \bar{u} - d \bar{d}) 
\quad  &\rightarrow& \;\displaystyle
{\cal C}_3=\frac{1}{\sqrt{2}} \,\lambda_3 \,,
 \\[0.4cm]
\qbq_{8}  : &\displaystyle
\frac{1}{\sqrt{6}} (u \bar{u} + d \bar{d} -2 s \bar{s}) 
\quad &\rightarrow&\;\displaystyle
{\cal C}_8=\frac{1}{\sqrt{2}} \,\lambda_8 \,,
 \\[0.4cm]
\qbq_{1} : & \displaystyle
\frac{1}{\sqrt{3}} (u \bar{u} + d \bar{d} +s \bar{s})
\quad  &\rightarrow& \; \displaystyle
{\cal C}_1= \frac{1}{\sqrt{n_f}}\, \mathbf{1}_f\,, 
\end{array}
\label{eq:ffactors} 
\end{equation}
where $\lambda_i$ are the usual $SU(3)$ Gell-Mann matrices and $\mathbf{1}$
is the $3\times 3$ unit matrix. For the flavor-singlet state, we use
the general notation \cite{Terentev81} in which the flavor content is expressed
in terms of $n_f$ which denotes the number of flavors
contained in $\qbq_{1}$ ($n_f=3$ in our case). This simplifies the 
comparison with the results for kernels to be found in the
literature. 

As usual \cite{LepageB80,BrodskyDFL86,ChernyakZ84} we define the
\da s in a frame where the meson moves along the
3-direction. Neglecting the meson's mass its momentum reads  
\begin{equation} 
p=[p^+,0,\mathbf{0}_{\perp}]
\, ,
\label{eq:p+}
\end{equation} 
where we use light-cone coordinates
 $v=[v^+, v^-, \mathbf{v}_\perp]$
with $v^\pm=(v^0\pm v^3)/\sqrt{2}$ for any four-vector $v$
\footnotemark[5].
We also introduce a light-like vector  
\begin{equation} 
n=[0,1,\mathbf{0}_{\perp}]
\, .
\label{eq:n-}
\end{equation} 
which defines the plus component of a vector, $v^+=n\cdot v$. 
The constituents of the meson, quarks or gluons, carry the fractions 
$u$ and $1-u$ of the light-cone plus components %
of the meson's momentum.
\begin{widetext}
The \da s are defined by Fourier transforms of hadronic matrix elements
\begin{eqnarray}
\Phi_{Pi} (u) &= & \frac{f_P^i}{2 \sqrt{2 N_c}} \, \phi_{Pi} (u)
  \nn
    \\
      & = &  -i \int \frac{dz^-}{2 \pi} e^{i(u-(1-u))p\cdot z}
        \;\left< 0 \left| \ov{\Psi}(-z) \;{\cal C}_i \; \frac{n\sla
        \gamma_5}{\sqrt{2 N_c}} \; \Omega \, \Psi(z) 
        \right| P\,(p)\right>
               \, ,
\label{eq:unPhiqgen}
\end{eqnarray}
and
\begin{eqnarray}
\Phi_{Pg}(u) &=& \frac{f_P^1}{2 \sqrt{2 N_c}} \,\phi_{Pg}(u)
\nn
   \\
 &=&  \frac{2 }{(n \cdot p)} \int \frac{dz^-}{2 \pi} 
      e^{i(u-(1-u))p\cdot z}
    \; \frac{n_{\mu} n_{\nu}}{\sqrt{N_c^2-1}} \; \left< 0 \left| 
      G^{\mu \alpha}(-z) \, \Omega \, \widetilde{G}_{\alpha}^{\: \: \: \nu}(z)
    \right| P\,(p) \right>
               \, ,
\label{eq:unPhiggen}
\end{eqnarray}
where $z=[0, z^-,\mathbf{0}_\perp]$. 
\footnotetext[5]{\\Different conventions for the light-cone components
are discussed in Ref. \cite{BrodskyPP98}.} \stepcounter{footnote}
\end{widetext}
Here, $\Psi$ denotes a quark field operator, 
$G^{\mu\nu}$ the gluon field strength tensor, 
and $\widetilde{G}^{\mu\nu}$ its dual 
\begin{equation} 
\widetilde{G}^{\mu \nu}=\frac{1}{2} \;
\epsilon^{\mu \nu \alpha \beta} \; G_{\alpha \beta}
\, .
\end{equation}
The quark and gluon operators in Eqs.\ (\ref{eq:unPhiqgen}),
(\ref{eq:unPhiggen}) are understood as color summed.
The path-ordered factor
\begin{equation}
 \Omega  =  
 \mbox{exp} \left\{ i g \int_{-1}^{1} ds A(z s)\cdot z \right\}
  \, ,
\label{eq:Omega}
\end{equation}
where $A$ is the gluon field, renders $\phi_{Pi}$ and $\phi_{Pg}$ gauge
invariant. The \da s in (\ref{eq:unPhiqgen}, \ref{eq:unPhiggen})
represent either the unrenormalized ones ($\phi^{ur}_{Pi,g}(u)$) if 
defined in terms of unrenormalized quark or gluonic composite 
operators or the renormalized one. In the latter case 
the \da s are scale dependent ($\phi_{Pi,g}(u,\mu^2)$). 
The \da s defined above satisfy the symmetry relations
\ba
\phi_{P1,8}(u,\mu^2)&=&\phi_{P1,8}(1-u,\mu^2) \, ,
\nn \\
\phi_{Pg}(u,\mu^2)&=&-\phi_{Pg}(1-u,\mu^2) \, .
\label{eq:DAsim}
\ea 

The definitions of the \da s \req{eq:unPhiqgen} and \req{eq:unPhiggen}
can be inverted to  
\ba
\lefteqn{\left< 0 \left| 
     \ov{\Psi}(-z) \;{\cal C}_i \; n\sla \gamma_5
           \; \Omega \, \Psi(z) \right| P\right>} 
\nn \\&  \qquad=&  i\, n \cdot p \, f_P^i \int_0^1 du \;
       e^{-i(2u-1)\, p\cdot z} \: \phi_{Pi}(u) \:
               \, , \qquad
\label{eq:melemq1u}
\ea
and
\ba
  \lefteqn{  \; n_{\mu} n_{\nu} \; \left< 0 \left| 
G^{\mu \alpha}(-z) \, \Omega \, \widetilde{G}_{\alpha}^{\: \: \: \nu}(z)
    \right| P \right> }
\nn \\ & = &   \frac12 (n \cdot p)^2\, \sqrt{C_F}\, f_P^1 \,
        \int_0^1 du \; e^{-i(2u -1)\, p \cdot z} \:   \phi_{Pg}(u) \:
                          \, .
\nn \\
\label{eq:melemg1u}
\ea

The projection of a collinear $q \bar{q}$ state onto a pseudoscalar
meson state is achieved by replacing the quark and antiquark spinors
(normalized  as $u^\dagger (p,\lambda) u(p,\lambda')=\sqrt{2} n\cdot p
\, \delta_{\lambda \lambda'}$)  by \cite{LepageB80}
\be
{\cal P}^{i,q}_{\alpha \beta, r s, k l}
= {\cal C}_{i, r s} \frac{\delta_{kl}}{\sqrt{N_c}}
 \left( \frac{\gamma_5 \not \! p}{\sqrt{2}} \right)_{\alpha \beta} 
\, , 
\label{eq:qq}
\ee
where $\alpha$ ($r$, $k$) and $\beta$ ($s$, $l$) represent Dirac
(flavor, color) labels of the quark and antiquark, respectively.
When calculating amplitudes,
the projector \req{eq:qq} leads to traces. The projector holds for
both incoming and outgoing states and corresponds to the definition of
the the quark \da s \req{eq:unPhiqgen}. 
It is to be used in calculations of hard-scattering amplitudes
which are to be convoluted with 
$f_{P}^i/(2 \sqrt{2 N_c}) \phi_{Pi}$ subsequently.

The form of the projection of a $gg$ state on a pseudoscalar
state with momentum $p$ can be deduced by noting that
the helicity zero combination of
transversal gluon polarization vectors $\eps^{\mu}$
can be written as \cite{DiehlFJK00}
\be
\begin{array}{l}
\eps^{\mu}(u p, \lambda) \eps^{\nu}((1-u) p, -\lambda)
- \eps^{\mu}(u p, -\lambda) \eps^{\nu}((1-u) p, \lambda) \\[0.15cm]
\qquad =   i\, \mbox{sign} (\lambda)\, \veps_{\perp}^{\mu \nu} 
\, ,
\end{array}
\label{eq:helzero}
\ee
where $\veps^{12}_\perp=-\veps^{21}_\perp=1$ while all other
components of the transverse polarization tensor are zero.
It can be expressed by
\begin{equation}
\veps_{\perp}^{\mu \nu} = \veps^{\mu \nu \alpha \beta}
\; \frac{n_{\alpha} p_{\beta}}{n\cdot p}
\, .
\label{eq:epsmunu}
\end{equation}
Instead of $n$ any other four vector can be used in \req{eq:epsmunu} 
that has a non-zero minus and a vanishing transverse component.
The projector of  an state of two incoming collinear gluons of
color $a$ and $b$ and Lorentz indices $\mu$ and $\nu$, 
associated with the momentum fractions $u$ and $(1-u)$,
respectively, onto a pseudoscalar meson state reads
\be
{\cal P}^{g}_{\mu \nu, a b} =\frac{i}{2}  \;
                 \frac{\delta_{ab}}{\sqrt{N_c^2-1}}
             \;  \frac{\veps_{\perp \mu \nu}}{u(1-u)} \,.
\label{eq:gg}
\ee
The complex conjugated expression is to be taken for an outcoming $gg$
state. The projector is to be used along with the \da{} 
$f_P^i/(2\sqrt{2N_c}) \phi_{Pg}$.   
The additional factor $[u(1-u)]^{-1}$ appearing as part of the
projector, is a consequence of the fact that in perturbative
calculations of reactions involving two-gluon Fock components, the 
potential $A$ of the gluon field occurs, while the gluon \da{} is 
defined in terms of the gluon field strength operator, see 
\req{eq:unPhiggen}. The conversion from a matrix element of field 
strength tensors \req{eq:melemg1u} to one of potentials is given by
\cite{Radyushkin96,Kogut} 
\begin{eqnarray}
\lefteqn{    \left< 0 \left| 
A^{\alpha}(-z) \, A^{\beta}(z)
    \right| P \right> }
\nn \\
   &\:= &  \frac{1}{4} \veps^{\alpha \beta}_\perp  
      \, \sqrt{C_F}\, f_P^1 \, 
    \int_0^1 du \; e^{-i(2 u -1)\, p \cdot z}
         \frac{\phi_{Pg}(u)}{u (1-u)}
\, . \qquad \quad
\label{eq:melemAu}
\end{eqnarray}
The gluonic projector \req{eq:gg} is obtained 
(up to the factor $[u(1-u)]^{-1}$ explained above)
by the coupling of two collinear gluons 
into a colorless pseudoscalar state.
In the context of mixing under evolution 
another normalization of it appears to be more appropriate,
see \req{eq:ggNEW}.
This normalization is accompanied by corresponding changes
in the gluon distribution amplitude $\phi_{Pg}$ and
the anomalous dimensions, as is discussed in detail in Sec. II.

For Levi-Civita tensor we use the convention
\begin{equation}
\veps^{0123}=-1 \, ,
\end{equation}
which leads to 
\begin{equation}
\textnormal{Tr} \left[ \gamma_5 \gamma^{\mu} \gamma^{\nu}
\gamma^{\alpha} \gamma^{\beta} \right]
=   4 i \veps^{\mu \nu \alpha \beta} \, 
\label{eq:Tr5}
\end{equation}
(with $\gamma_5=i \gamma^0 \gamma^1 \gamma^2 \gamma^3$).

\section{The $P\gamma$ transition form factor -\protect\\
details of the calculation}
\label{app:calc}
In this appendix, we provide some details of the calculation
of the evolution kernels and the hard scattering amplitude for
the flavor-singlet contribution to the $P\gamma$ transition
form factor. These quantities can, in principle, be taken from 
the literature 
(see, e.g. \cite{Terentev81,BaierG81} and \cite{DVCS}%
\footnote{In Ref. \cite{DVCS} the NLO corrections to the 
deeply virtual Compton amplitude 
$\gamma^* p \to \gamma^* p$ have been calculated.
In the limiting case of zero skewness the Compton amplitude
is related to our process by crossing.}
) but the conventions 
and notations differ. However, since it is imperative to use a 
consistent set of conventions for the hard scattering amplitude 
and the \da s, we recalculate them. In doing so we follow 
closely Ref.\ \cite{MelicNP01}. Dimensional regularization in
$D=4-2\epsilon$ dimensions is used to regularize UV and 
collinear singularities which appear when calculating the 
one-loop diagrams. According to \cite{MelicNP01}, the 
$\gamma_5$ problem, i.e, the ambiguity which enters the 
calculation due to the presence of one $\gamma_5$ matrix and 
the use of dimensional regularization method, is resolved 
by matching the results for the hard-scattering part with the 
results for the perturbatively calculable part of the \da{}, 
since the physical form factor is free of ambiguity.
We employ the $\overline{\rm MS}$ coupling constant renormalization
along the same lines as in \cite{MelicNP01}.
We note in passing, that as long as the singularities are not fully removed
from the amplitudes, the following relations 
are to be used
for the change of the scale of the coupling constant
\begin{equation}
   \als{(\mu^2)} = \left( \frac{\muR}{\mu^2} \right)^\eps 
                    \, \als(\muR) \, 
        \big[ 1 + {\cal O} (\als)\big] \, 
\label{eq:chalphaS}
\end{equation}
and for the $\beta$ function
\begin{equation}
\beta(\alpha_s(\mu^2),\eps) = \mu^2 \frac{\partial }{\partial
\mu^2}
        \alpha_s(\mu^2)
         =
   - \eps \, \alpha_s(\mu^2)
   - \frac{\alpha_s^2(\mu^2)}{4 \pi} \beta_0
          \, .
\label{eq:beta}
\end{equation}
\begin{figure*}
   \centerline{\includegraphics[width=11cm]{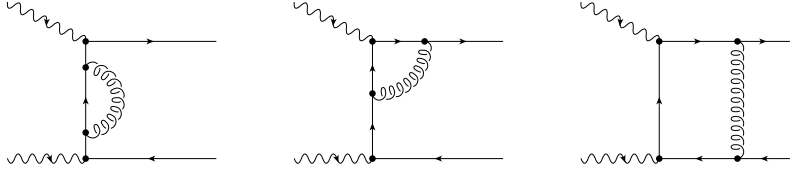}}
\caption{Sample NLO Feynman diagrams contributing to the
  $\gamma^* \gamma \rightarrow \qbq$ amplitude.}
\label{f:NLOq}
\end{figure*}
\begin{figure*}
   \centerline{\includegraphics[width=13cm]{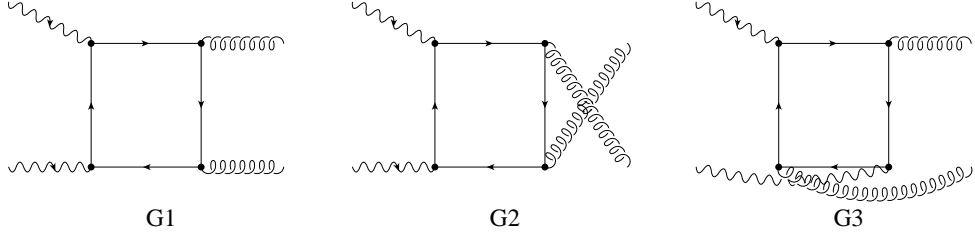}} 
 \caption{Distinct one-loop Feynman diagrams contributing to the
  $\gamma^* \gamma \rightarrow gg$ amplitude.
  Other contributing diagrams are obtained from these by
  reversing the direction of the fermion flow in the loops.}
\label{f:eHSA}
\end{figure*}
The usual renormalization group coefficient is given by
\begin{equation}
\beta_0=\frac{11}{3} N_c - \frac{2}{3} n_f \, .
\label{eq:beta0}
\end{equation}

\subsection{Amplitudes}
\label{s:T}
The amplitude $\gamma\gamma\to \qbq$
denoted by $T_{\qbq}$ 
(examples of contributing Feynman diagrams are depicted
in  Fig.\ \ref{f:NLOq})
has the structure already 
quoted in (\ref{eq:expT}) where 
\begin{eqnarray}
T_{q \bar{q}}^{(0)}(u) & = & \frac{1}{1-u} + \frac{1}{u} 
\, ,
\nn\\[0.3em]
T_{q \bar{q}}^{(1)}(u) & = & 
        \frac{-1}{\eps} {\cal A}_{col, q \bar{q}}^{(1)}(u)
      + {\cal A}_{q \bar{q}}^{(1)}(u)  \, .
\label{eq:Tqq}
\end{eqnarray}
The functions ${\cal A}$ read
\begin{eqnarray}
{\cal A}_{col, \qbq}^{(1)}(u) &=& \frac{1}{1-u} 
          \Big[3 + 2 \ln(1-u) \Big] + (u \rightarrow 1-u)
                                          \, , \nn\\[0.3cm]
{\cal A}_{\;\qbq\;}^{(1)}(u) &=&
            \frac{1}{1-u} \left[ -9 - \frac{1-u}{u} \ln(1-u) 
           + \ln^2(1-u) \right] 
\nn \\ & &+ (u \rightarrow 1-u)
      \, . 
\label{eq:aqq}
\end{eqnarray}
In obtaining the above results the projector \req{eq:qq} is employed.
The results for the flavor-octet and singlet
cases differ only in the flavor factors 
(see \req{eq:c1} and \req{eq:c8}).

Next, we calculate the amplitude $T_{gg}$ for the subprocess 
$\gamma^* \gamma \rightarrow gg$. The appropriate gluonic
projector is the complex conjugate of (\ref{eq:gg}). 
For the case of the transition form factor we can work in a Breit 
frame where the momentum of the real photon, $q_2$,
is proportional to the vector $n$ from Eq.\ \req{eq:n-},
and can therefore be employed in \req{eq:epsmunu}. 
There are 6 one-loop diagrams that contribute to this subprocess 
amplitude. Three representative diagrams ($G1$, $G2$, $G3$) 
are shown in Fig.\ 
\ref{f:eHSA}. The other three reduce to the first three ones by 
reversing the direction of the fermion flow in the loop. 
Moreover, it is easy to see that
\begin{equation}
T_{G2}= - T_{G1}(u \rightarrow 1-u) \, .
\end{equation}
Thus, one has only to calculate the contributions from the  
diagrams $G1$ and $G3$. 

The complete unrenormalized NLO contribution is the sum of 
individual contributions in which, expectedly, the UV singularities 
cancel. The hard-scattering amplitude $T_{gg}$ has the structure 
quoted in (\ref{eq:expT}) where $T^{(1)}_{gg}$ is given by 
\begin{equation}
T_{gg}^{(1)}(u) =
  \frac{-1}{\eps} {\cal A}_{col,gg}^{(1)}(u) +
                  {\cal A}_{gg}^{(1)}(u) 
 \, .
\label{eq:Tgg1}
\end{equation}
The functions ${\cal A}$ read%
\footnote{Making use of the crossing
relations, it can be shown that the functions
\req{eq:aqq} and \req{eq:agg} are in agreement with 
the coefficient functions for the Compton amplitude
quoted in \cite{DVCS}.}
\begin{eqnarray}
{\cal A}_{col,gg}^{(1)}(u) & = &
2 \left[ \frac{1}{u^2} \ln (1-u) - (u \rightarrow 1-u) \right]
\, ,
 \nonumber \\[0.3cm]
{\cal A}_{\: \: gg\: \:}^{(1)}(u)& = & 
\frac{2}{u (1-u)} \left[
\left( 3 - \frac{2}{u} \right) \ln (1-u) 
\right. \nn \\ & & \left.
+ \frac{1-u}{2 u} \ln^2 (1-u) 
- (u \rightarrow 1-u)
\right]
\, . 
 \nn \\ & & 
\label{eq:agg}
\end{eqnarray}

\subsection{Kernels}
\label{s:Vij}
For the calculation of the renormalization matrix $Z$, respective 
$V^{(1)}$ in (\ref{eq:Zexp}) we utilize the method proposed in 
\cite{MelicNP01,Katz85} of saturating the mesonic state by its 
valence Fock components (\ref{eq:osbasis}) which leads to
\be
\Phi_P^{ur}(u)= -i \tilde{\phi}(u,v) \otimes \left(
\begin{array}[c]{c}
\langle \qbq_1; v|P \rangle \\[0.2em]
\langle gg; v|P \rangle 
\end{array}
\right)   
\, .
\label{eq:Phiurtotilde}
\ee
The elements of the matrix $\tilde{\phi}$ are defined as in 
\req{eq:unPhiqgen} and \req{eq:unPhiggen} with the replacement
of $|P\rangle$ by $|q \overline{q}_1\rangle$ and $|gg\rangle$.
They are thus perturbatively calculable
and determine the matrix $Z$.

The calculation of the matrix element $Z_{qq}$ proceeds along 
the same lines as indicated for the flavor-octet case in Ref.\
\cite{MelicNP01} and the contributing diagrams are displayed there.
The respective kernel $V_{qq}$ reads
\begin{eqnarray}
   V_{qq}(u,v) &=&
   2\, C_F\,
  \left\{ \frac{u}{v} 
    \left[ 1 + \frac{1}{v-u} \right] 
         \Theta(v-u) 
  \right. \nn \\ & & \left.
                    + 
        \bigg( \begin{array}{c} u \rightarrow 1-u \\
                                v \rightarrow 1-v
               \end{array} \bigg) \right\}_+
\, ,
\label{eq:Vqq} 
\end{eqnarray}
where the usual plus distribution
is defined as
\begin{equation}
   \Big\{ F(u,v) \Big\}_+ \equiv F(u,v) 
        - \delta(u-v) \int_0^1 dz \, F(z,v)
              \, .
\label{eq:F+}
\end{equation}
This result also holds for the flavor-octet case.

\begin{figure}
\centerline{\includegraphics[width=6.7cm]{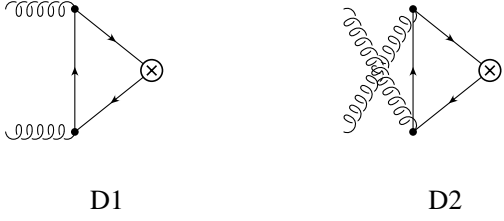}} 
\caption{LO Feynman diagrams that contribute to $\tilde{\phi}_{qg}$.
The crossed circle denotes the vertex of 
$\langle 0 |
\overline{\Psi}(-z)\, {\cal C}_1\,n\sla\gamma_5/\sqrt{2 N_c} 
\Psi(z)$.}
\label{f:DAqg}
\end{figure}
We proceed to the evaluation of $Z_{qg}$, or rather $V_{qg}$.
According to the definition of the $\qbq_1$ \da{}, the
matrix element that is of interest here, is given by
($z=[0,z^-,\tr{0}]$)
\ba
 \tilde{\phi}_{qg}(u)
  &=&  
   \int \frac{dz^-}{2 \pi} 
     e^{i(2u-1)p\cdot z}
\nn \\ & & \times
    \; \left< 0 \left| 
     \ov{\Psi}(-z) \;{\cal C}_1 \; \frac{n\sla \gamma_5}{\sqrt{2 N_c}}
           \; \Omega \, \Psi(z) 
    \right| gg \right>
               \, ,
\nn \\
\label{eq:tphiqg}
\ea
The relevant Feynman diagrams for the calculation of 
$\tilde{\phi}_{qg}$ are depicted in Fig. \ref{f:DAqg}. The 
$\qbq$ vertex, $\otimes$, is of the form \cite{MelicNP01,Katz85}
\be
         {\cal C}_{1}
    \frac{\mathbf{1}_{c}}{\sqrt{N_c}}\:
         \frac{n\sla\gamma_5}{2\sqrt{2}}\,
         \delta(u\, n\cdot p - n\cdot k)\,,
\label{eq:vertex}
\ee
where $k$ represents the momentum of the quark entering 
the circle.
The vertex \req{eq:vertex} occurs also 
in the calculation of the $\tilde{\phi}_{qq}$
where the LO contribution is obtained
by contracting the vertex just with the 
$\qbq$ projector \req{eq:qq} and, hence, one obtains 
$\tilde{\phi}_{qq}(u,v)=\delta(u-v)$ as it should be
(see \req{eq:Zexp}).

Due to the presence of only one $\gamma_5$ matrix,
we are confronted with the $\gamma_5$ problem,
as in the calculation of $T_{q \bar{q}}$. 
When using the naive $\gamma_5$ scheme, in which 
the $\gamma_5$ matrix retains 
its anticommuting properties in $D$ dimensions,
we obtain three different results depending on the
position of $\gamma_5$ inside the trace:\\
\begin{eqnarray}
\lefteqn{\tilde{\phi}_{qg,D1}(u,v)}
\nn \\  & = &
                       - \sqrt{n_f C_F}\, \frac{\als}{4 \pi} \,
           \left\{ \frac{(4 \pi)^2}{i} \,
                \left[ \mu^{2 \eps} \int \frac{d^D l}{(2 \pi)^D}
              \frac{1}{(l^2 + i \eta )^2} \right] \right\} 
\nn \\ & & \times
                   \left[ \frac{u }{v^2} \Theta(v-u) -
                         \frac{(1-u)}{(1-v)^2} \Theta(u-v) 
                                \right. \nonumber \\ 
          & & \left. + \frac{2 \eps}{1-\eps/2} \delta
                   \left( \frac{u}{v^2(1-v)} \Theta(v-u) 
\right. \right. \nn \\ & & \left. \left. +
                   \frac{(1-u)}{v(1-v)^2} \Theta(u-v) \right)
                                \right] 
\, ,
\label{eq:phiD1res}
\end{eqnarray}
where 
\begin{equation}
\delta \in \{ - (2 v -1), -1, 1 \} \, .
\label{eq:delta}
\end{equation}
The loop integral can be worked out analytically%
\footnote{
The treatment of the integral in Eq.\ \req{eq:phiD1res}
was explained in detail in \cite{MelicNP01}. The crucial point 
is to retain a distinction between UV and collinear singularities.} 
and we refer to \cite{MelicNP01} for the result.

One can easily see that 
\begin{equation}
 \tilde{\phi}_{qg,D2}(u,v) = - \tilde{\phi}_{qg,D1}(u, 1-v)
 \, ,
\label{eq:phiD2}
\end{equation}
and finally
\begin{equation}
 \tilde{\phi}_{qg}(u,v) = \tilde{\phi}_{qg,D1}(u, v)
           - \tilde{\phi}_{qg,D1}(u, 1-v)
 \, .
\label{eq:phiqg}
\end{equation}
The kernel $V_{qg}$ is a residue of the UV singularity
embodied in the loop integral appearing in \req{eq:phiD1res} 
and, hence, is related to the term multiplying the integral 
in \req{eq:phiD1res}.
Since the term proportional to $\delta$ is finite
($\sim \eps \, (1/\eps)$), it does not contribute to $V_{qg}$.
Moreover, since $\tilde{\phi}_{qg}$ being antisymmetric
under the replacement of $v$ by $1-v$, is to be convoluted
with the matrix element $\left<gg | P \right>$ 
(see \req{eq:Phiurtotilde}),
which has the same symmetry properties as the full gluon
\da{} (see \req{eq:DAsim}), 
one can replace $\tilde{\phi}_{qg}$ 
by $\tilde{\phi'}_{qg}(u,v) = 2 \tilde{\phi}_{qg,D1}(u,v)$
in order to obtain a more compact representation of the kernel
\begin{equation}
V_{qg}(u,v)  = 
- 2 \sqrt{n_f C_F}
\left\{
\frac{u }{v^2} \Theta(v-u) 
  - \left( \begin{array}{c}
                       u \rightarrow 1-u \\
                       v \rightarrow 1-v 
                        \end{array} \right)
                       \right\}
\, .
\label{eq:Vqg}
\end{equation}

The set of LO evolution kernels is completed by 
\ba
V_{gq}(u,v) &=& 2 \sqrt{n_f C_F} \left\{ \frac{u^2}{v} \Theta(v-u)
                       - \left( \begin{array}{c}
                                            u \rightarrow 1-u \\
                                            v \rightarrow 1-v 
                                             \end{array} \right)
                                            \right\}
\, ,\nn \\ & &
\label{eq:Vgq}   \\[0.4em]
{V}_{gg}(u,v) &=& 2 N_c
\left\{ \frac{u}{v} \left[\left(\frac{\Theta(v-u)}{v-u}\right)_+ 
             + \frac{2 u -1}{v}\Theta(v-u) \right]
\right. \nn \\ & & \left.
          + \left( \begin{array}{c}
                    u \rightarrow 1-u \\
                    v \rightarrow 1-v 
                    \end{array} \right) \right\}
          + \,\beta_0\, \delta(u-v)\,.
\label{eq:Vgg} 
\ea
Since, except of the normalization, there is general agreement in
the literature on these kernels, see e.g.\ \cite{Terentev81,BaierG81},
we quote them without giving any detail of their calculation. 
Finally, we comment on an alternative definition 
of the gluon distribution amplitude
which one occasionally encounters in the literature.
In that definition the factor $[u(1-u)]^{-1}$ is included in
$\phi_{Pg}$ instead in the $gg$ projector \req{eq:gg}.  
The results for $T_{gg}$ (\ref{eq:Tgg1}), (\ref{eq:agg}) 
will, hence, be multiplied
by $u(1-u)$, while the kernels take the form
\begin{equation}
\begin{array}{ccc}
V_{qg} \rightarrow V_{qg} \; v (1-v) \, , &\qquad&
\displaystyle V_{gq} \rightarrow \frac{V_{gq}}{u (1-u)}  \, ,\\[0.2cm]
\multicolumn{3}{c}{ \displaystyle
V_{gg} \rightarrow V_{gg} \; \frac{v (1-v)}{u (1-u)}
\, . }
\end{array}
\label{eq:altV}
\end{equation}
The result for the transition form factor, 
as for any other physical quantity,
is, obviously,
invariant under the redefinition of the gluon \da{}.

\section{Some properties of the evolution kernel} 
\label{app:prop}
It is easy to verify that the evolution kernels (\ref{eq:Vqq}) 
and (\ref{eq:Vqg})-(\ref{eq:Vgg}) satisfy the symmetry
relations 
\begin{eqnarray}
 v\phantom{2} (1-v)\phantom{2} \; V_{qq}(u,v)  &=&  
                 u\phantom{2} (1-u)\phantom{2} \; V_{qq}(v,u) \, , 
                                              \nn\\[0.2em]
v^2 (1-v)^2 \; V_{gg}(u,v) &=&  u^2 (1-u)^2 \: V_{gg}(v,u) \, ,  
                                              \nn\\[0.2em]
v^2 (1-v)^2 \; V_{qg}(u,v)  &=&  
                  u\phantom{2} (1-u)\phantom{2} \: V_{gq}(v,u) \, .
\end{eqnarray}

\begin{widetext}
The kernels $V_{ij}$, convoluted with the weighted Gegenbauer
polynomials $C_n^m$ of order $m=3/2, 5/2$, result in
\begin{eqnarray}
V_{qq}(u,v) \, \otimes \, v\phantom{2} (1-v)\phantom{2} 
                                             \, C_n^{\,3/2}(2 v-1)
             &=& \gamma^{qq}_n  \; u\phantom{2} (1-u)\phantom{2} 
                                               \, C_n^{\,3/2}(2v-1)
                                            \,, \nn\\[0.3em]
V_{qg}(u,v) \, \otimes \, v^2 (1-v)^2 \; C_{n-1}^{5/2}(2 v-1)
             &=& \gamma^{qg}_n  \; u\phantom{2}(1-u)\phantom{2} 
                                              \; C_n^{\,3/2}(2v-1)
                                             \,,\nn \\[0.3em]
V_{gq}(u,v) \, \otimes \, v\phantom{2} (1-v)\phantom{2} 
                                              \,\, C_{n}^{3/2}(2 v-1)
          &=& \gamma^{gq}_n  \; u^2(1-u)^2 \, C_{n-1}^{5/2}(2v-1)
                                               \,,\nn \\[0.3cm]
V_{gg}(u,v) \, \otimes \, v^2 (1-v)^2 \; C_{n-1}^{5/2}(2 v-1)
          &=& \gamma^{gg}_n  \; u^2(1-u)^2 \,\, C_{n-1}^{5/2}(2v-1)
\, .
\label{eq:VC}
\end{eqnarray}
\end{widetext}
The factors on the right hand side of \req{eq:VC} multiplying the 
Gegenbauer polynomials are the anomalous dimensions. 
The results
quoted for them in (\ref{eq:andim}) can be read off from \req{eq:VC}
(for a detailed discussion see \cite{BaierG81}).
Finally, we mention that the off-diagonal anomalous
dimensions in \req{eq:andim} satisfy the relation
\begin{equation}
   \frac{\gamma_n^{qg}}{\gamma_n^{gq}} =
   \frac{N_{n-1}^{\,5/2}}{N_n^{\,3/2}} 
\, ,
\label{eq:gammaN}
\end{equation}
where 
\begin{eqnarray}
N_n^{3/2} &=& \frac{(n+1)(n+2)}{4 (2n+3)} \, ,\\
N_{n-1}^{5/2} &=& \frac{n(n+3)}{36} N_n^{3/2} \, ,
\, 
\end{eqnarray}
represent the
normalization constants of the corresponding Gegenbauer polynomials
\begin{eqnarray}
\int_0^1 du\, u (1-u)\, C_n^{\,3/2}(2 u-1) C_m^{3/2}(2 u-1)
                             &=& N_n^{\,3/2}\, \delta_{nm} \, , \nn\\
\int_0^1 du\, u^2 (1-u)^2\, C_n^{\,5/2}(2 u-1) C_m^{5/2}(2 u-1)
                             &=& N_n^{\,5/2}\, \delta_{nm} \, .
\nn \\  &&
\end{eqnarray}
Throughout the paper we investigate only the LO behavior
of the evolution kernels and corresponding anomalous dimensions.
Beyond leading order, the relations corresponding to 
\req{eq:VC} and \req{eq:gammaN} get modified 
due to mixing of conformal operators starting at NLO
 (see, for example, \cite{BelitskyMS99}).

\end{appendix}

\end{document}